\documentclass[12pt]{iopart}

\usepackage{graphicx}
\usepackage{latexsym}
\usepackage{iopams, amsopn, amstext, wasysym}
\usepackage{colonequals}
\usepackage{dcolumn}
\usepackage{setspace}
\usepackage{subfigure}
\usepackage{bm}
\usepackage{rotating}
\usepackage{braket}
\usepackage{units}
\usepackage{multirow}
\usepackage{xspace}
\usepackage[usenames,dvipsnames]{color}

\usepackage[bookmarksnumbered, bookmarksopen, breaklinks, colorlinks]{hyperref}

\usepackage{ulem}
\newcommand{\ihope}{\texttt{ihope}}
\newcommand{\subrows}[2][c]{%
  \begin{tabular}[#1]{@{}c@{}}#2\end{tabular}}

\newcommand{\define}{\colonequals}
\newcommand{\eob}{\texttt{EOBNRv2}\xspace}
\newcommand{\imr}{\texttt{PhenomB}\xspace}
\newcommand{\imrns}{\texttt{PhenomB$_\mathtt{\chi=0}$}\xspace}

\def\IL{\relax{\rm I\kern-.18em L}}

\newcommand{\Ms}{\ensuremath{\mathrm{M}_{\odot}}}
\newcommand{\msun}{\ensuremath{\mathrm{M}_\odot}}

\newcommand{\Overlap}{\Braket}

\def\ltsima{$\; \buildrel < \over \sim \;$}
\def\simlt{\lower.5ex\hbox{\ltsima}}
\def\gtsima{$\; \buildrel > \over \sim \;$}
\def\simgt{\lower.5ex\hbox{\gtsima}}

\makeatletter
\let\protect\relax
{\catcode`\|=\active
  \xdef\InnerProduct{\protect\expandafter\noexpand\csname InnerProduct
\endcsname}
  \expandafter\gdef\csname InnerProduct \endcsname#1{%
    \begingroup
    \ifx\SavedDoubleVert\relax
    \let\SavedDoubleVert\|\let\|\IpDoubleVert
    \fi
    \mathcode`\|32768\let|\IPVert
    \left({#1}\right)
    \endgroup
  }
}
\def\IPVert{\@ifnextchar|{\|\@gobble}
     {\egroup\,\mid@vertical\,\bgroup}}
\def\IPDoubleVert{\egroup\,\mid@dblvertical\,\bgroup}
\let\SavedDoubleVert\relax
\def\midvert{\egroup\mid\bgroup}
\def\SetVert{\@ifnextchar|{\|\@gobble}
    {\egroup\;\mid@vertical\;\bgroup}}
\def\SetDoubleVert{\egroup\;\mid@dblvertical\;\bgroup}
\def\mid@vertical{\mskip1mu\vrule\mskip1mu}
\def\mid@dblvertical{\mskip1mu\vrule\mskip2.5mu\vrule\mskip1mu}
\makeatother

\begin{document}

\title[The NINJA-2 project]{The NINJA-2 project: Detecting and characterizing 
gravitational waveforms modelled using numerical binary black hole 
simulations.}

\author{J.~Aasi$^{1}$, 
B.~P.~Abbott$^{1}$, 
R.~Abbott$^{1}$, 
T.~Abbott$^{2}$, 
M.~R.~Abernathy$^{1}$, 
T.~Accadia$^{3}$, 
F.~Acernese$^{4,5}$, 
K.~Ackley$^{6}$, 
C.~Adams$^{7}$, 
T.~Adams$^{8}$
% ...
}

\address{$^{1}$LIGO - California Institute of Technology, Pasadena, CA  91125, 
USA }
\address{$^{2}$Louisiana State University, Baton Rouge, LA  70803, USA }
\address{$^{3}$Laboratoire d'Annecy-le-Vieux de Physique des Particules (LAPP), 
Universit\'e de Savoie, CNRS/IN2P3, F-74941 Annecy-le-Vieux, France}
\address{$^{4}$INFN, Sezione di Napoli, Complesso Universitario di Monte 
S.Angelo, I-80126 Napoli, Italy}
\address{$^{5}$Universit\`a di Salerno, Fisciano, I-84084 Salerno, Italy}
\address{$^{6}$University of Florida, Gainesville, FL  32611, USA }
\address{$^{7}$LIGO - Livingston Observatory, Livingston, LA  70754, USA }
\address{$^{8}$Cardiff University, Cardiff, CF24 3AA, United Kingdom }
% ...

\begin{abstract}
The Numerical INJection Analysis (NINJA) project is a collaborative
effort between members of the numerical relativity and
gravitational-wave astrophysics communities.  The purpose of NINJA is
to study the ability to detect gravitational waves emitted from merging 
binary black holes and recover their parameters with next-generation 
gravitational-wave observatories. We report here on the results of the second 
NINJA project, 
NINJA-2, which employs 60 complete binary black hole hybrid waveforms 
consisting of a numerical 
portion modelling the late inspiral, merger, and ringdown stitched to a 
post-Newtonian portion modelling the early inspiral. In a ``blind injection 
challenge'' similar to that conducted in recent LIGO and Virgo science
runs, we added 7 hybrid waveforms to two months of data recolored to 
predictions of
Advanced LIGO and Advanced Virgo sensitivity curves during their first 
observing runs. The resulting data 
was analyzed by gravitational-wave detection algorithms and 6 of the 
waveforms were recovered with false alarm rates smaller than 1 in a 
thousand years. Parameter estimation algorithms were run on each of 
these waveforms to explore the ability to constrain the masses, 
component angular momenta and 
sky position of these waveforms. We find that the strong degeneracy between 
the mass ratio and the black holes' angular momenta will make it difficult to
precisely estimate these parameters with Advanced LIGO and Advanced Virgo. We 
also perform a large-scale monte-carlo study to assess the ability to recover 
each of the 60 hybrid waveforms with early Advanced LIGO and Advanced Virgo 
sensitivity curves. 
Our results predict that early Advanced LIGO and Advanced Virgo will have a 
volume-weighted 
average sensitive distance of 300Mpc (1Gpc) for $10M_{\odot}+10M_{\odot}$ 
($50M_{\odot}+50M_{\odot}$) binary black hole coalescences. We 
demonstrate that neglecting the component angular momenta in the waveform 
models used in matched-filtering will result in a reduction in sensitivity for 
systems with large component angular momenta. 
This reduction is estimated to be up to $\sim15\%$ for 
$50M_{\odot}+50M_{\odot}$ 
binary black hole coalescences with almost maximal angular momenta aligned with 
the orbit when using early Advanced LIGO and Advanced Virgo sensitivity curves.
\end{abstract}
  
\maketitle

\section{Introduction}
\label{sec:introduction}
A network of second-generation laser interferometric
gravitational-wave (GW) observatories is presently under construction.
The US-based Advanced Laser Interferometer Gravitational Wave 
Observatory (aLIGO)~\cite{Harry:2010zz} is expected to have its
initial observing run in 2015 utilizing observatories in Hanford,
Washington and Livingston, Louisiana (denoted ``H'' and ``L'', respectively). 
aLIGO will then work towards reaching design
sensitivity, expected in 2018-20~\cite{Aasi:2013wya}. The
French-Italian Advanced Virgo (AdV) observatory~\cite{Accadia:2011zzc,aVIRGO}
(denoted ``V'') is expected to follow shortly after the
aLIGO instruments. The cryogenically cooled KAGRA
observatory~\cite{Kuroda:2010zzb,Somiya:2011np} and a India-based aLIGO
facility~\cite{LIGODCC:M1100296,Unnikrishnan:2013qwa} are due to begin 
operations around
2020, providing a 5-site network to explore the gravitational-wave sky
in detail.

These second-generation observatories will have an order of magnitude
increase in sensitivity over their first generation counterparts and
will be sensitive to a broader range of gravitational-wave
frequencies~\cite{Harry:2010zz,aVIRGO,Somiya:2011np}. One of the
primary observational targets for this global network is the inspiral, merger 
and ringdown of a binary system containing two black 
holes~\cite{thorne.k:1987}. 
With aLIGO and AdV operating at their final design sensitivities
it is expected that 0.4 - 1000 binary black hole (BBH) coalescences
will be observed per year of operation~\cite{Abadie:2010cf}.  Directly
observing the collision of two black holes will allow
gravitational-wave astronomers to understand the physics of black-hole
spacetimes and to explore the strong-field conditions of the
theory of general relativity~\cite{Sathyaprakash:2009xs}.

Exploring the underlying mass and spin distributions of stellar-mass black
holes can tell us a great deal about the end stages of massive-star evolution.
The mass measurements of compact objects made to date suggest a gap between the
most massive neutron stars ($\lesssim 3~\msun$)~\cite{Kalogera:1996ci} and the
least massive black holes ($\gtrsim 5~\msun$)~\cite{Bailyn:1997xt}.  It is
still an open question as to whether this gap is real and the result of
formation mechanisms, or simply due to observational biases~\cite{Farr:2010tu}.
Whether or not ground-based detectors will be able to distinguish between
these regions in mass space is of great interest. Furthermore, from the
distributions of black hole spin magnitudes and tilts (orientation of the spin
relative to the orbital angular momentum), more can be learned about supernovae
kicks and compact binary formation environments.  Stellar-mass black hole spin
measurements are currently done by modelling either the accretion disk's thermal
continuum X-ray spectrum, or the profile of the broadened Fe K$\alpha$
line~\cite{McClintock:2011zq}.  Both methods are fundamentally based on
assumptions about the location of the inner edge of the accretion disk, and
also depend sensitively on very complicated physical models of the disk and its
emission.  Gravitational waves will provide an entirely new method of
measuring black hole spin which does not require the complicated modeling of
accretion disk physics. 

Of the stellar mass black hole angular momenta (spin) measurements made to 
date, half are found to have a magnitude
$a\gtrsim0.8$~\cite{McClintock:2011zq}. With BBH observations, aLIGO and AdV 
will be able to provide independent measurements of the black hole spin 
magnitudes. Therefore, it will be interesting
to evaluate how well aLIGO and AdV will be able to constrain the magnitude of 
the black holes' component spins.
The direction of the compact objects' angular momenta is also
of interest, with particular implications for formation
mechanisms~\cite{Belczynski:2007xg}.  Measuring systems with component
spins misaligned with the orbital angular momentum is outside of the
scope of this project. However, this study does include systems with
component spins that are both aligned and anti-aligned with the
orbital angular momenta, and we will evaluate the ability of aLIGO and AdV to
distinguish such systems from one another.

The standard technique for observing BBH mergers
involves matched-filtering data taken from gravitational-wave observatories 
against ``template'' waveforms that should closely match potential 
astrophysical signals~\cite{Wainstein:1962,Wainstein:1968,Allen:2005fk}. The
observable BBH waveform includes the signal from the inspiral of the
two black holes, as well as their merger and the resulting black
hole's ringdown. Search templates must include all of these
features~\cite{Buonanno:2009zt,Brown:2012nn}.  As an alternative to
matched-filter searches, a number of algorithms exist to perform
searches for \emph{unmodelled} gravitational-wave signals
\cite{Klimenko:2008fu,Searle:2007uv,Sutton:2009gi}. These algorithms
do not require accurate knowledge of the waveforms to make
observations, but are not as sensitive as matched-filter searches in
cases where the waveform models are well understood.

Theoretical models of the inspiral, merger and ringdown of BBH systems
are necessary to produce template banks for matched-filter searches
and to use as model signals to test both matched-filter and unmodelled
searches.  The inspiral portion of the waveform can be modeled by
analytic post-Newtonian (PN) calculations
\cite{Blanchet:2006zz,Buonanno:2009zt}, while numerical solutions of
the General Relativity field equations are required to accurately
model the final orbits and merger.  Prior to breakthroughs in
numerical relativity (NR) in
2005~\cite{Pretorius:2005gq,Campanelli:2005dd, Baker:2005vv}, template
banks and search pipeline tests used only inspiral waveforms.  Since
2007, NR waveforms have been used to calibrate analytical waveform
models~\cite{Buonanno:2007pf,Damour:2009kr,Pan:2011gk,Taracchini:2012ig,
Ajith:2009bn,Santamaria:2010yb,Damour:2012ky,Taracchini:2013rva}. Some
of the analytical waveforms have been already employed in search
pipelines~\cite{Aasi:2012rja}.  However, there exists another
useful and valuable avenue of communication between numerical
relativists and gravitational-wave astronomers.  As NR pushes into new
regions of parameter space the waveforms can be used directly to test
searches employing previously-calibrated templates, and the degree to
which these searches prove to be insufficient can motivate both new
template models and additional simulations.

The Numerical INJection Analysis (NINJA) project was created in 2008. 
The project uses recent 
advances in numerical relativity~(\cite{Centrella:2010mx} and references
therein) to test analysis 
pipelines by adding numerically-modelled, physically-realistic signals to 
detector noise and attempting to recover these signals with search pipelines.
The first NINJA project (NINJA-1)~\cite{Aylott:2009ya} utilized a total of
23 numerical waveforms, which were injected into Gaussian noise colored with
the frequency sensitivity of initial LIGO and Virgo. These data were
analyzed by nine data-analysis groups using both search and
parameter-estimation algorithms~\cite{Aylott:2009ya}.  

However, there were four limitations to
the NINJA-1 analysis. First, due to the computational cost of NR simulations, 
most waveforms 
included only $\sim$10 orbits before merger. Therefore the waveforms were 
too short to inject over an astrophysically
interesting mass range without introducing artifacts into the data. The lowest
mass binary considered in NINJA-1 had a total mass of 35$\Ms$, whereas
the mass of black holes could extend below
$5\Ms$~\cite{Farr:2010tu,Ozel:2010su}.
Second, the waveforms were only inspected for obvious,
pathological errors and no cross-checks were performed between the
submitted waveforms. It was therefore difficult to assess the
physical fidelity of the results.  
Third, the NINJA-1 data set
contained stationary noise with the simulated signals already injected
into the data. Since the data set lacked the non-Gaussian noise transients 
present 
in real detector data, it was not possible to fully explore the response of the
algorithms in a real search scenario. Finally, the data set contained only 126 
simulated signals,
this precluded detailed statistical studies of the effectiveness of
search and parameter estimation algorithms. Despite these limitations,
the NINJA-1 project lead to a framework within which to perform injection 
studies using waveforms as calculated by the full nonlinear general theory of 
relativity, established guidelines for such studies (in particular a 
well-defined format
for the exchange of NR waveforms~\cite{Brown:2007jx}), and clarified where 
further work was needed.

This lead to the initiation of the second NINJA project (NINJA-2), whose goals 
were to build and improve upon NINJA-1 and perform a systematic test of the 
efficiency of data-analysis pipelines in preparation for the Advanced detector 
era. A set of 60 NR waveforms were submitted by 8 numerical relativity groups 
for the NINJA-2 project~\cite{Ajith:2012az}. These waveforms conform 
to a set of length and accuracy requirements, and
are attached to PN inspiral signals to produce hybrid PN-NR waveforms that can 
be injected over the full range of physically relevant total binary masses.
The construction and verification of these waveforms is described in a previous 
paper~\cite{Ajith:2012az}, and summarized here in 
section~\ref{sec:waveforms}. In the Numerical-Relativity and 
Analytical-Relativity collaboration, a project complementary to the NINJA 
collaboration, 22 new NR waveforms were produced and rigorously analysed. These 
newly produced NR waveforms were compared to the most recent calibrated 
analytical models, finding that the loss of event rates due to modeling is 
below 3\%~\cite{Hinder:2013oqa}. In this paper we study the ability of the 
search algorithms used in
the last of the Initial LIGO and Virgo science runs to observe
numerically-modelled BBH waveforms from the set of 60 waveforms
submitted to the NINJA-2 project. This is done using data taken during LIGO's
sixth and Virgo's second science runs and recoloring that data to the
sensitivities expected from early observation runs of aLIGO and AdV.

There are a wide range of search and parameter estimation algorithms available
within the GW astronomy community: those that were used in past analyses of
detector data, old algorithms that have been updated and re-tuned following the
experience gained in those analyses, plus many new algorithms under
development. For both practical reasons, and to mark a clear point in the
development and refinement of these methods, this work employed \emph{only}
search and parameter estimation algorithms that were approved and used in the
last initial-LIGO and Virgo science runs, \emph{without any additional tuning
or 
modifications}~\cite{Klimenko:2008fu,Babak:2012zx,Aasi:2012rja,
Abadie:2012rq}. By doing this we aim to provide a benchmark against which
future algorithms can be compared.

A set of 7 numerical relativity waveforms, with
masses ranging from 14.4$M_{\odot}$ to 124$M_{\odot}$ were added into
the recolored data as an unbiased test of the process through which candidate 
events are identified for BBH waveforms. This 
data was distributed to analysts who
knew that such ``blind injections'' were present but had no
information about the number, parameters or temporal location of these
waveforms. This was similar to blind injection tests conducted by the
LIGO and Virgo collaborations in their latest science runs
\cite{Colaboration:2011np}. Using a search for unmodelled gravitational wave 
transients we found that one of these signals was recovered, with an estimated 
false alarm rate of 1 every 47 years. The
remaining 6 signals were consistent with background. Using a
matched-filtered algorithm with a bank of BBH IMR waveforms, which were 
\emph{not} calibrated against the NR signals used in NINJA-2, 6 of the
signals were recovered with more significance than all background
events. This allowed upper limits on the false alarm rate ranging
between 1 every 5000 years and 1 every 40000 years to be placed on
each blind injection. The remaining signal was not recovered due to
having a low network signal-to-noise ratio and possessing a large
anti-aligned spin, which was not modelled in the bank of waveforms used in the 
search.

Parameter estimation algorithms have come a long way since the first NINJA
project. Previously these analyses were unable to estimate the parameters of
high mass systems accurately due to the use of inspiral-only models on data
with little measurable inspiral. We show that these tools are now capable of
reliably providing parameter estimates for both low and high mass systems.
For all but one injection the masses and spins of the black holes were recovered
within the estimated 95\% credible regions.  The remaining injection suffered
small systematic biases due to non-Gaussian features present in the noise, the
modeling of which is an ongoing endeavor.
We find that strong intrinsic degeneracies between the masses and black hole 
spins \cite{Baird:2012cu,Hannam:2013uu} make it difficult to constrain the 
masses well, for 3 of the signals the 
presence of a neutron star cannot be ruled out. We also investigate the ability 
to constrain the sky localization of the various signals and demonstrate how 
even low power non-Gaussian noise transients in the data can effect the 
recovery of the intrinsic parameters of BBH systems.

We use large sets of known waveforms to assess the 
efficiency of the matched-filter BBH search algorithm as a 
function of the mass and angular momenta of the component black holes. 
These are the first such studies that have been done using real data recolored 
to second-generation noise curves, which include the non-Gaussian features that 
will be present in the data taken with Advanced LIGO and Advanced Virgo. 
As these results were obtained using the search pipelines and techniques that
were deployed in the final observing runs of Initial LIGO and Initial Virgo, 
they can therefore provide a benchmark against which improvements to the search 
techniques can be compared and assessed. 

In our large-scale simulation studies we find 
evidence that incorporating search waveforms including the effects of spin will 
increase the efficiency of searches. The results shown here can be used, in the 
future, to 
compare with results of search pipelines including the effects of component 
spins that are aligned with the orbital angular momentum, which are currently 
under development~\cite{Brown:2012qf,Ajith:2012mn,Harry:2013tca}. We also 
assess 
the efficiency of the matched-filter 
BBH search algorithm to recover waveforms generated by different 
groups with the same parameters. We find that the 
efficiency of the matched-filter BBH search algorithm to recover 
different waveforms, generated by different groups, but with identical physical 
parameters, is indistinguishable up to statistical errors.

In this work we have not scaled observed masses and distances to account for 
cosmological effects, which will be important especially for high-mass binary 
black hole collisions. Therefore any masses and distances quoted should be 
interpreted as \emph{observed} masses and \emph{luminosity} distances.

The paper is organized as follows:
In section~\ref{sec:waveforms} we briefly summarize the waveform
catalogue described more fully in~\cite{Ajith:2012az}.
Section~\ref{sec:noise} describes the LIGO/Virgo data used and the
processing that was done to make it resemble anticipated
advanced-detector noise.  Section~\ref{sec:parameters} describes how
the parameters for the signals were chosen and reports the values that
were selected.  Section~\ref{sec:pipelines} describes the detection
algorithms that were run on the data set and section~\ref{sec:searches} reports
their results.  Section~\ref{sec:param_estimation} describes
the parameter estimation results. 
Section~\ref{sec:sensitivity} describes the results of 
a high-statistics analysis aimed at quantifying the sensitivity of the 
detection searches to the different hybrid 
waveforms. We conclude in section~\ref{sec:conclusions} with a discussion of 
how well the various algorithms performed, and implications for the
Advanced detector era.

\section{PN-NR Hybrid Waveforms}
\label{sec:waveforms}

The NINJA-2 waveform catalog contains 60 PN-NR hybrid waveforms that 
were contributed by eight numerical relativity groups.  This catalog and the 
procedures used to validate it are described in detail in
~\cite{Ajith:2012az}. We briefly summarize the NINJA-2 catalog 
here.

\begin{figure} 
\centerline{\includegraphics[width=0.95\linewidth]
{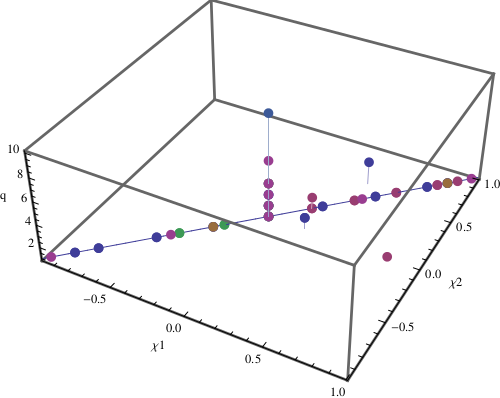}}
  \caption[Parameters of the NINJA-2 submissions]{
  \label{fig:ParameterSpace}
Mass ratio $q$ and dimensionless spins $\chi_i$ of the NINJA-2 hybrid
waveform submissions. Reproduced from \cite{Ajith:2012az}.
}
\end{figure}

Each waveform in the NINJA-2 waveform catalog consists of a PN 
portion modelling the early
inspiral, stitched to a numerical portion modelling the late inspiral,
merger and ringdown.  This ensures accurate modelling of the late
portions of the waveform while simultaneously ensuring that waveforms
are long enough to be scaled to masses as low as $10 M_\odot$ without
starting abruptly within the sensitive frequency band of the
detectors.  We require that for the NR portion of the waveform the
amplitude be accurate to within 5\% 
and the phase (as a function of gravitational-wave frequency)
have an accumulated uncertainty over the inspiral, merger and ringdown
of no more than 0.5 rad.  Since we do not have access to exact
waveforms we define ``accuracy'' by convergence of the numerical
waveforms as resolution and waveform-extraction radius are increased.
We also require at least five orbits of numerical data in order to
ensure robust blending with the PN portion.  No requirements were
placed on the hybridization itself, although it is known that
hybridization can introduce significant errors
\cite{MacDonald:2011ne, Santamaria:2010yb, Ohme:2011rm}.
It was
decided to limit NINJA-2 to systems without eccentricity, and with
black-hole spins parallel or anti-parallel
to the orbital angular momentum.  This
last condition avoids precession, which we do for two reasons; (i)
precession greatly complicates waveform phenomonology
and we prefer to first tackle a simpler subset which still maintains
the main features of binary evolution and merger; and (ii) at the
start of NINJA-2 the precessing-binary parameter space had been
sampled by only a handful of numerical simulations.  Waveforms were
submitted in the format described in \cite{Brown:2007jx}, and data was
provided as strain decomposed into spherical harmonics of weight $-2$.
Groups were encouraged to submit modes beyond $(l,m)=(2,\pm 2)$
and many 
did so.  However the techniques to validate these higher modes
are a current research topic.  In order not to delay the NINJA-2
project it was decided to validate only the $(2,\pm 2)$ modes in
~\cite{Ajith:2012az} and employ only these modes for the first
NINJA-2 analysis. Different groups employed different codes, as well
as different methods for solving initial conditions, dealing with
singularities, evolving Einstein's equations, and extracting
gravitational-wave information.  In addition different PN
approximants and different hybridization methods were used by
different groups in constructing the full hybrid waveforms. 
It was found that the
dominant source of disagreement between submissions was in the PN
portion, and in particular overlaps between submissions were greater
than 0.97 over the range of
masses, including regions sensitive to differences in hybridization
techniques.  See
\cite{Ajith:2012az} for details.

The parameter space for aligned-spin BBH systems is four dimensional;
the masses and spin magnitudes of each of the two holes.  However, in
the absence of matter Einstein's equations possess a mass invariance,
and a solution obtained by numerical relativity or other method may be
trivially rescaled to any total mass.  We therefore eliminate total
mass from the parameter space of submissions leaving the ratio of the
two masses, denoted $q$, and the dimensionless spins denoted
$\chi_{1,2}$ which must lie between $-1 < \chi_{1,2} < 1$.

Tables
\ref{tab:ninja2_submissions1} and \ref{tab:ninja2_submissions2} give a
summary of the submissions for systems where the masses of the two
black holes are equal and unequal, respectively. The first column of 
Tables~\ref{tab:ninja2_submissions1}
and~\ref{tab:ninja2_submissions2} gives a label for each waveform, to
ease referring to them in later sections.  These labels of the
form ``{\tt G2+20+20\_T4}'' are constructed as follows: The first letter
represents the group submitting the numerical simulation:
\begin{itemize}
\item[{\bf F}:] The numerical relativity group at Florida Atlantic University, 
also using the BAM 
code~\cite{Tichy:2010qa,Brugmann:2008zz,Marronetti:2007ya,Bruegmann:2003aw}.
\item[{\bf G}:] The Georgia Tech group using 
MayaKranc~\cite{Healy:2008js,Healy:2009ir,Bode:2009mt,Herrmann:2007ex,
Healy:2009zm,Bode:2011tq,Hinder:2007qu}
\item[{\bf J}:] The BAM (Jena) code, as used by the Cardiff-Jena-Palma-Vienna
  
collaboration~\cite{Husa:2007hp,Hannam:2007wf,Ajith:2009bn,Hannam:2010ec,
Brugmann:2008zz,Hannam:2007ik}
\item[{\bf L}:] The Lean Code, developed by Ulrich 
Sperhake~\cite{Sperhake:2006cy,Sperhake:2007gu}.
\item[{\bf Ll}:] The Llama code, used by the AEI group and the Palma-Caltech 
groups~\cite{Pollney:2010hs,Reisswig:2009rx,Pollney:2009yz}
\item[{\bf R}:] The group from Rochester Institute of Technology, using the
  LazEv code~\cite{Campanelli:2005dd,Lousto:2010tb,Lousto:2010qx,Nakano:2011pb}.
\item[{\bf S}:] The SXS collaboration using the SpEC 
code~\cite{Pfeiffer:2002wt,Scheel:2006gg,Lovelace:2011nu,Szilagyi:2009qz,
Lovelace:2010ne,Scheel:2008rj,SpECWebsite,Boyle:2007ft,Lindblom:2005qh,
Boyle:2009vi}.
\item[{\bf U}:] The group from The University of Illinois~\cite{Etienne:2008re}.
\end{itemize}
Immediately after this letter follows the mass-ratio $q=m_1/m_2$,
where the black holes are labeled such that $q\ge 1$.  Subsequently
are the components of the initial dimensionless spin along the orbital
angular momentum, multiplied by 100 (e.g. `+20' corresponds to $\hat
L\cdot \vec S_1 /m_1^2=0.2$) of the more massive and the less massive
black hole.  The label closes with the Taylor-approximant being used
for the PN portion of the waveform, with ``T1'' and ``T4'' representing
TaylorT1 and TaylorT4, respectively.  The Georgia Tech group submitted
four pairs of simulations where each pair simulates systems with identical 
physical parameters, stitched to the same PN approximant. These 
waveforms are not identical however as each simulation within a pair has a 
different number of NR cycles and was generated at a different resolution.  
These are distinguished by appending ``\_1'' and ``\_2'' to the label.

Each NR group verified that their waveforms met the minimum NINJA-2
requirements as described above.  The minimum-five-orbits requirement
was easily verified by inspection, and the amplitude and phase
uncertainties were estimated by convergence tests with respect to
numerical resolution and waveform-extraction radius.  The full catalog
was then verified by the NINJA-2 collaboration.  Submissions were inspected in
the time and frequency domains to identify any obvious problems caused
by hybridization or integration from the Newman-Penrose curvature
scalar $\psi_4$ to strain.  Where multiple simulations were available
for the same physical parameters these simulations were compared using
the matched-filter \emph{overlap}.  The inner product between two real
waveforms $s_1(t)$ and $s_2(t)$ is defined as
\begin{equation}
\label{eq:InnerProduct}
     \InnerProduct{s_1|s_2} 
 = 4\, \Re \int_{0}^\infty df\,
   \frac
     {\tilde{s}_1(f) \tilde{s}_2^\star(f)}
     {S_n(f)}
\end{equation}
where $\tilde{x}$ denotes the Fourier transform of $x$ and $S_n(f)$ is
the power spectral density, which was taken to
be the target sensitivity for the first advanced-detector runs,
referred to as the ``early aLIGO'' PSD. This is described in more detail in 
section \ref{sec:noise}.

The overlap is then
obtained by normalization and maximization over relative time and
phase shifts, $\Delta t$ and $\Delta \phi$.
\begin{equation}
  \label{eq:OverlapDefinition}
  \Overlap{s_1|s_2} \define 
  \max_{\Delta t, \Delta \phi} \frac{\InnerProduct{s_1|s_2}}{
    \sqrt{\InnerProduct{s_1|s_1} \InnerProduct{s_2|s_2}}}.
\end{equation}
The investigations in \cite{Ajith:2012az} demonstrated that
the submitted waveforms met the requirements as outlined above and in
addition were consistent with each other to the extent expected.  We
therefore conclude that these submissions model real gravitational
waves with sufficient accuracy to quantitatively determine how
data-analysis pipelines will respond to signals in next-generation
gravitational-wave observatories.

The NINJA-2 waveforms cover the 3-dimensional aligned-spin parameter
space rather unevenly, as indicated in figure~\ref{fig:ParameterSpace}.
The configurations available fall predominantly into two 1-dimensional
subspaces: (i) Binaries of varying mass-ratio, but with non-spinning
black holes.  (ii) Binaries of black holes with equal-mass and
equal-spin, and with varying spin-magnitude. Future studies, with additional 
waveforms covering the gaps that are clearly evident in 
figure~\ref{fig:ParameterSpace} 
and waveforms including 
precession~\cite{Campanelli:2008nk,Mroue:2013xna,Hinder:2013oqa}, 
would be useful to more fully understand the 
response of search codes across the parameter space, and would help to better 
tune analytical waveform models including inspiral, merger and ringdown phases.
\begin{table}
\caption[Submissions to NINJA-2]{
\label{tab:ninja2_submissions1}
Summary of the contributions to the NINJA-2 waveform catalog with $m_1
= m_2$.  Given are an identifying label, described in
section~\ref{sec:waveforms}, mass-ratio $q=m_1/m_2$ which is always
$1$ for these simulations, magnitude of the dimensionless spins
$\chi_i=S_i/m_i^2$, orbital eccentricity $e$, frequency range of
hybridization in $M\omega$, the number of numerical cycles from the
middle of the hybridization region through the peak amplitude, and the
post-Newtonian Taylor-approximant(s) used for hybridization.
}
  \begin{indented}
    \item[]\begin{tabular}{@{}rrrcrrcc}
      \ms\bhline\ms
      Label & $q$ & $\chi_{1}$ & $\chi_{2}$ & $1000e$   & $100\,M\omega$ & \# NR 
& pN \\
      &    &     &            &            & hyb.range & cycles & Approx \\
      \mr
S1-95-95\_T1 & 1.0  &  -0.95  &  -0.95  &  1.00  &   3.3 -- 4.1  &  18.42  &  T1 
\\
J1-85-85\_T1 & 1.0  &  -0.85  &  -0.85  &  2.50  &   4.1 -- 4.7  &  12.09  & 
T1\\
J1-85-85\_T4 & & & & & & & T4 \\
J1-75-75\_T1 & 1.0  &  -0.75  &  -0.75  &  1.60  &   4.1 -- 4.7  &  13.42  & 
T1\\
J1-75-75\_T4 & & & & & & & T4 \\
J1-50-50\_T1 & 1.0  &  -0.50  &  -0.50  &  2.90  &   4.3 -- 4.7  &  15.12  & 
T1\\
J1-50-50\_T4 & & & & & & & T4 \\
S1-44-44\_T4 & 1.0  &  -0.44  &  -0.44  &  0.04  &   4.3 -- 5.3  &  13.47  &  T4 
\\
Ll1-40-40\_T1 & 1.0  &  -0.40  &  -0.40  &    &   6.1 -- 8.0  &  6.42  & T1\\
Ll1-40-40\_T4 & & & & & & & T4 \\
J1-25-25\_T1 & 1.0  &  -0.25  &  -0.25  &  2.50  &   4.5 -- 5.0  &  15.15  & 
T1\\
J1-25-25\_T4 & & & & & & & T4 \\
Ll1-20-20\_T1 & 1.0  &  -0.20  &  -0.20  &    &   5.7 -- 7.8  &  8.16  & T1\\
Ll1-20-20\_T4 & & & & & & & T4 \\
J1+00+00\_T1 & 1.0  &  0.00  &  0.00  &  1.80  &   4.6 -- 5.1  &  15.72  & T1\\
J1+00+00\_T4 & & & & & & & T4 \\
G1+00+00\_T4 &   &    &    &  3.00  &  5.5 -- 7.5  &  9.77  &  T4 \\
Ll1+00+00\_F2 &   &    &    &    &  5.7 -- 9.4  &  8.30  &  F2 \\
S1+00+00\_T4 &   &    &    &  0.05  &  3.6 -- 4.5  &  22.98  &  T4 \\
G1+20+20\_T4\_1 & 1.0  &  0.20  &  0.20  &  10.00  &   6.0 -- 7.5  &   6.77  &  
T4 \\ 
G1+20+20\_T4\_2 &      &        &        &   6.00  &   5.5 -- 7.5  &  10.96  &  
T4 \\ 
J1+25+25\_T1 & 1.0  &  0.25  &  0.25  &  6.10  &   4.6 -- 5.0  &  18.00  & T1\\
J1+25+25\_T4 & & & & & & & T4 \\
G1+40+40\_T4\_1 & 1.0  &  0.40  &  0.40  &  10.00  &   5.9 -- 7.5  &   7.70  &  
T4 \\ 
G1+40+40\_T4\_2 &      &        &        &   6.00  &   5.5 -- 7.5  &  12.02  &  
T4 \\ 
Ll1+40+40\_T1 &   &    &    &    &  7.8 -- 8.6  &  6.54  & T1\\
Ll1+40+40\_T4 & & & & & & & T4 \\
S1+44+44\_T4 & 1.0  &  0.44  &  0.44  &  0.02  &   4.1 -- 5.0  &  22.39  &  T4 
\\
J1+50+50\_T1 & 1.0  &  0.50  &  0.50  &  6.10  &   5.2 -- 5.9  &  15.71  & T1\\
J1+50+50\_T4 & & & & & & & T4 \\
G1+60+60\_T4\_1 & 1.0  &  0.60  &  0.60  &  12.00  &   6.0 -- 7.5  & 8.56  &  T4 
\\   
G1+60+60\_T4\_2 &      &        &        &   5.00  &   5.5 -- 7.5  &  13.21  &  
T4 \\ 
J1+75+75\_T1 & 1.0  &  0.75  &  0.75  &  6.00  &   6.0 -- 7.0  &  14.03  & T1\\
J1+75+75\_T4 & & & & & & & T4 \\
G1+80+00\_T4 & 1.0  &  0.80  &  0.00  &  13.00  &   5.5 -- 7.5  &  12.26  &  T4 
\\ 
G1+80+80\_T4\_1 & 1.0  &  0.80  &  0.80  & 14.00  &   5.9 -- 7.5  &  9.57  &  T4 
\\ 
G1+80+80\_T4\_2 &      &        &        &  6.70  &   5.5 -- 7.5  & 14.25  &  T4 
\\
J1+85+85\_T1 & 1.0  &  0.85  &  0.85  &  5.00  &   5.9 -- 6.9  &  15.36  & T1\\
J1+85+85\_T4 & & & & & & & T4 \\
U1+85+85\_T1 &   &    &    &  20.00  &  5.9 -- 7.0  &  15.02  &  T1 \\
G1+90+90\_T4 & 1.0  &  0.90  &  0.90  &  3.00  &   5.8 -- 7.5  &  15.05  &  T4 
\\
S1+97+97\_T4 & 1.0  &  0.97  &  0.97  &  0.60  &   3.2 -- 4.3  &  38.40  &  T4 
\\
      \ms\bhline\ms
    \end{tabular}
  \end{indented}
\end{table}

\begin{table}
\caption[Submissions to NINJA-2]{
\label{tab:ninja2_submissions2}
Summary of the contributions to the NINJA-2 waveform catalog with $m_1
> m_2$.  Given are an identifying label, described in 
section~\ref{sec:waveforms}, 
mass-ratio $q=m_1/m_2$
magnitude of the dimensionless spins $\chi_i=S_i/m_i^2$, orbital
eccentricity $e$, frequency range of hybridization in $M\omega$, the
number of numerical cycles from the middle of the hybridization region
through the peak amplitude, and the post-Newtonian Taylor-approximant(s)
used for hybridization.
}
  \begin{indented}
    \item[]\begin{tabular}{@{}rrrcrrcc}
      \ms\bhline\ms
      Label & $q$ & $\chi_{1}$ & $\chi_{2}$ & $1000e$   & $100\,M\omega$ & \# NR 
& pN \\
      &    &     &            &            & hyb.range & cycles & Approx \\
      \mr
J2+00+00\_T1 & 2.0  &  0.00  &  0.00  &  2.30  &   6.3 -- 7.8  &  8.31  & T1\\
J2+00+00\_T4 & & & & & & & T4 \\
G2+00+00\_T4 &   &    &    &  2.50  &  5.5 -- 7.5  &  10.42  &  T4 \\
Ll2+00+00\_F2 &   &    &    &    &  6.3 -- 9.4  &  7.47  &  F2 \\
S2+00+00\_T2 &   &    &    &  0.03  &  3.8 -- 4.7  &  22.34  &  T2 \\
G2+20+20\_T4 & 2.0  &  0.20  &  0.20  &  10.00  &   5.6 -- 7.5  &  11.50  &  T4 
\\
J2+25+00\_T1 & 2.0  &  0.25  &  0.00  &  2.00  &   5.0 -- 5.6  &  15.93  & T1\\
J2+25+00\_T4 & & & & & & & T4 \\
J3+00+00\_T1 & 3.0  &  0.00  &  0.00  &  1.60  &   6.0 -- 7.1  &  10.61  & T1\\
J3+00+00\_T4 & & & & & & & T4 \\
S3+00+00\_T2 &   &    &    &  0.02  &  4.1 -- 5.2  &  21.80  &  T2 \\
F3+60+40\_T4 & 3.0  &  0.60  &  0.40  &  1.00  &   5.0 -- 5.6  &  18.89  &  T4 
\\
J4+00+00\_T1 & 4.0  &  0.00  &  0.00  &  2.60  &   5.9 -- 6.8  &  12.38  & T1\\
J4+00+00\_T4 & & & & & & & T4 \\
L4+00+00\_T1 &   &    &    &  5.00  &  5.1 -- 5.5  &  17.33  &  T1 \\
S4+00+00\_T2 &   &    &    &  0.03  &  4.4 -- 5.5  &  21.67  &  T2 \\
S6+00+00\_T1 & 6.0  &  0.00  &  0.00  &  0.04  &   4.1 -- 4.6  &  33.77  &  T1 
\\
R10+00+00\_T4 & 10.0  &  0.00  &  0.00  &  0.40  &   7.3 -- 7.4  &  14.44  &  T4 
\\
      \ms\bhline\ms
    \end{tabular}
  \end{indented}
\end{table}

\section{Modified Detector Noise}
\label{sec:noise}

In this section we describe the techniques used in this work to emulate data 
that will be taken by second generation gravitational wave observatories. This 
was accomplished by recoloring data taken from the initial LIGO and Virgo 
instruments to predicted 2015 -- 2016 sensitivities. Recoloring initial LIGO 
and 
Virgo data allows the non-Gaussianity and non-stationarity of that data to be 
maintained.

\begin{figure}
\centering
\includegraphics[width=0.495\textwidth]
{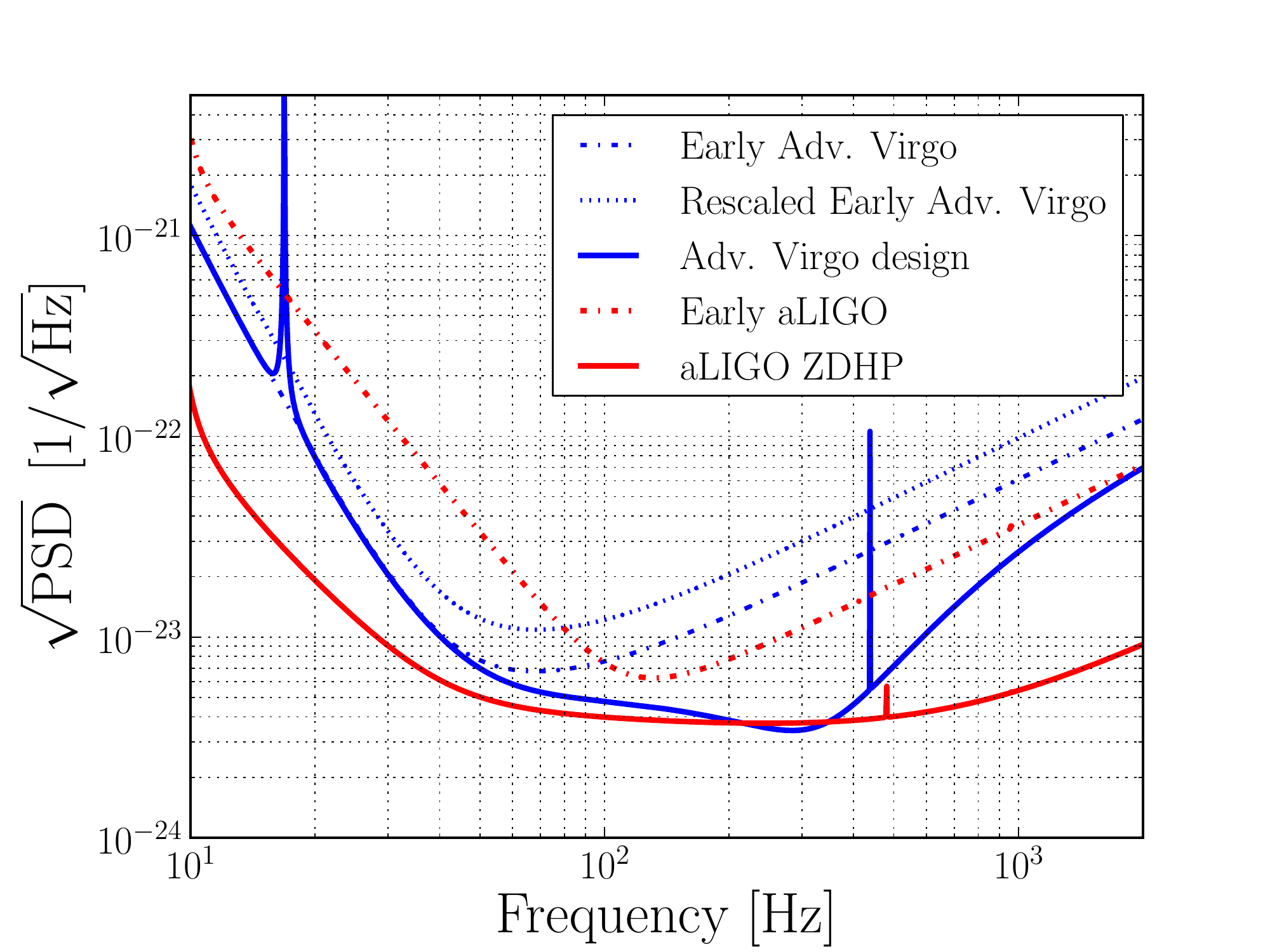}
\includegraphics[width=0.495\textwidth]
{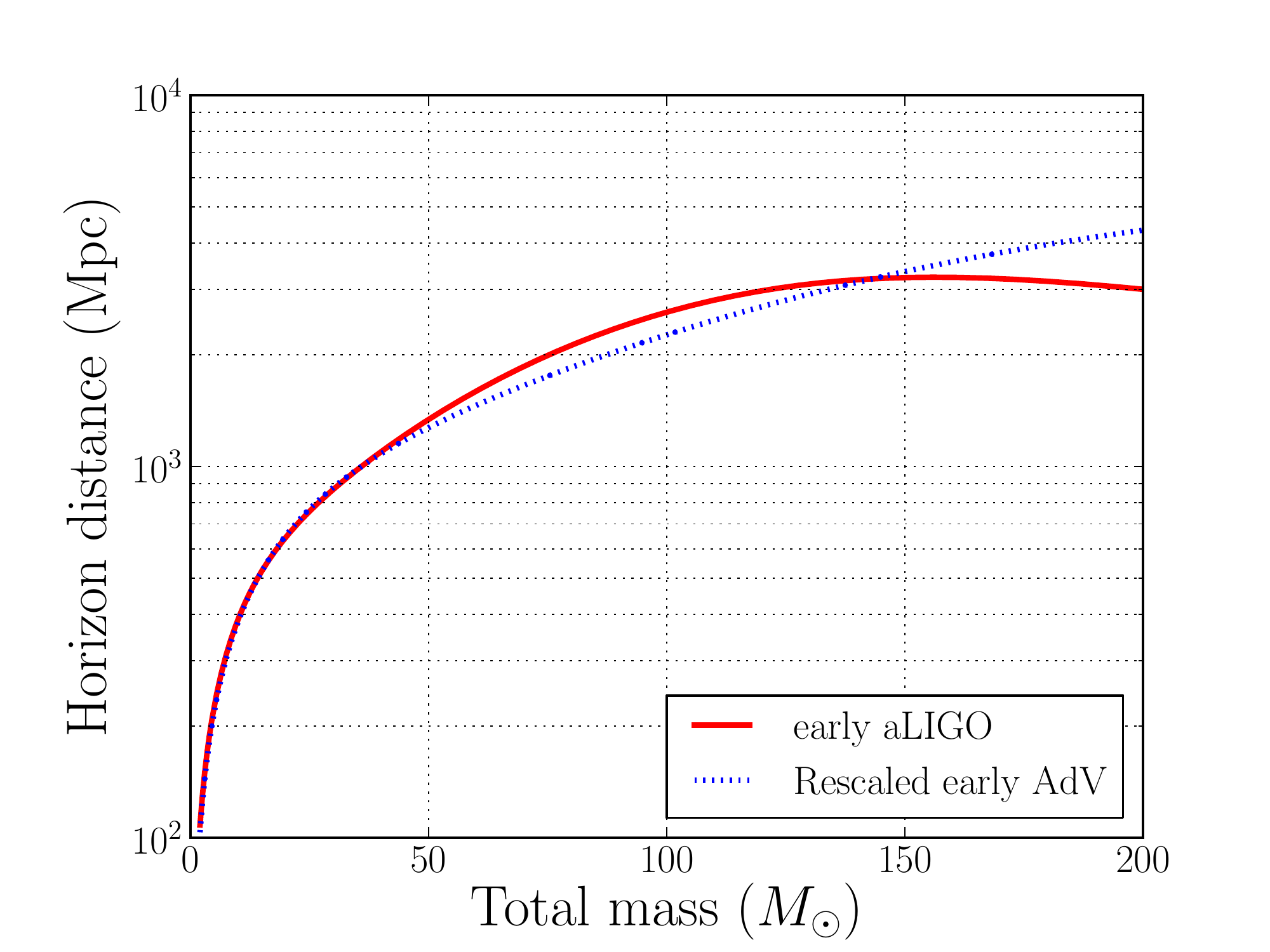}
\caption{\label{fig:NOISE_design_spectra}
Left: predicted sensitivity curves for aLIGO and
AdV. Shown are both the design curves and predicted 2015 -- 2016
\emph{early} sensitivity curves. Also shown is the early AdV noise curve 
rescaled such that the horizon distance for a (10 $M_{\odot}$, 10 $M_{\odot}$) 
binary system is equal to that obtained with the early aLIGO noise 
curve. Right: Horizon distance as a function of observed total mass for the 
early aLIGO and rescaled early AdV sensitivity curves. This plot is 
made considering only equal mass, non-spinning systems and calculated using the 
\eob ~\cite{Pan:2011gk} waveform approximant. Results in this paper are 
generated from the early aLIGO noise curve and the rescaled early AdV curve.}
\end{figure}

The predicted sensitivity curves of the advanced detectors as a function of
time can be found in the living document
\cite{Aasi:2013wya}. For this work we are interested in the
sensitivity of the advanced detectors in 2015 -- 2016 and used a previous 
prediction of the sensitivity curves for this time period as given in
\cite{LV_early_noisecurves} and shown in the left panel of figure 
\ref{fig:NOISE_design_spectra}.
These curves were used as the updated predictions given in 
\cite{Aasi:2013wya} were not available when we began this study.
We refer to the 2015 -- 2016 predicted noise curves as the \emph{early} 
sensitivity curves.
It is clear from the figure that the predicted
sensitivity of early AdV is significantly greater than 
that of the early aLIGO curve, when using the predictions given in 
\cite{LV_early_noisecurves}. In the right panel of figure 
\ref{fig:NOISE_design_spectra} we
show the distance at which optimally oriented, optimally located,
non-spinning, equal mass binaries would be detected with a
signal-to-noise ratio (SNR) of 8 using both noise curves.
This is commonly referred to as the \emph{horizon distance}. The early AdV 
noise curve was rescaled
by a factor of 1.61 so that the sensitive distance for a
(10 $M_{\odot}$, 10 $M_{\odot}$) binary merger would be equal to the early aLIGO
noise curve. This rescaling was found to better reflect the updated predicted
sensitivities presented in \cite{Aasi:2013wya}. The results in this 
paper are generated using the early aLIGO and rescaled early AdV sensitivity 
curves.

As with the initial science runs, we expect data taken from
these detectors, in the absence of gravitational-wave signals, to be neither
Gaussian nor stationary. It is important that search pipelines demonstrate
an ability to deal with these features. To simulate data with 
advanced detector sensitivities and with
realistic non-Gaussian and non-stationary features, we chose to use
data recorded by initial LIGO and Virgo and recolor that data to the
predicted early sensitivity curves of aLIGO and AdV. The data we
chose to recolor was data taken during LIGO's sixth science run and Virgo's
second science run.

The procedure for producing such \emph{recolored data} was accomplished in the
following steps, which were conducted separately for the two LIGO detectors and
Virgo.

\begin{itemize}
 \item Identify a two-month duration of initial detector data to be 
recolored
 \item Measure the power spectral density (PSD) for each distinct section 
of science mode data using the PSD estimation routines in the \texttt{lal} 
software package \cite{LAL}.
 \item Calculate an \emph{average PSD} over the two month period by taking, for
every discrete value of frequency recorded in the PSDs, the median value over
each of the PSDs in the set.
 \item Remove any line features from the resulting PSD and from the predicted
early noise curves. This is done because it is difficult to remove or introduce 
line features from the data without introducing unwanted artifacts. Therefore 
it is simpler to remove line features in the PSDs, which will have the effect 
of preserving the line features of the original data into the recolored 
data.
 \item For each frequency bin, record the median value of the PSD over each 
section of science mode
 \item Take the ratio of the median PSD and the predicted early advanced
detector noise curve.
 This is the reweighting to be used when recoloring.
 \item Using the time domain filtering abilities of the \texttt{gstlal}
software package \cite{GSTLAL}, recolor the data using this reweighting factor.
\end{itemize}

\begin{figure}
\centering
\includegraphics[width=0.495\textwidth]
{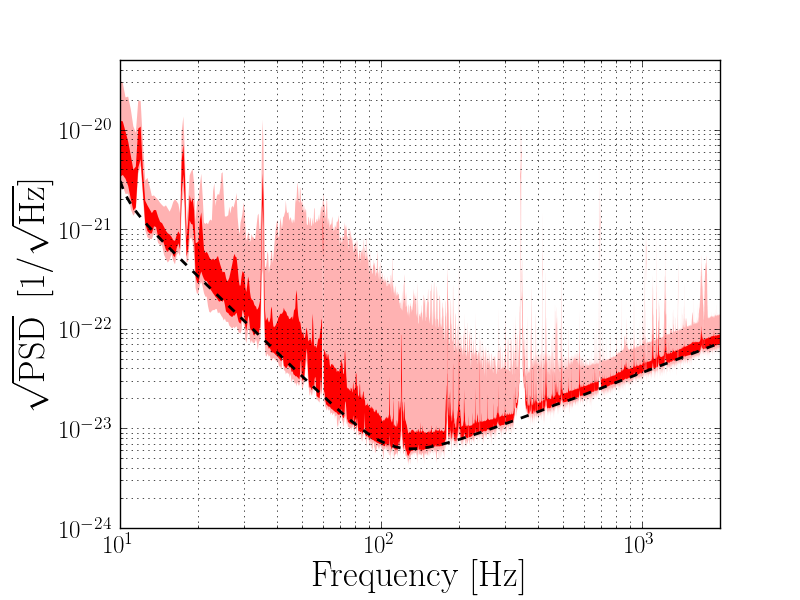}
\includegraphics[width=0.495\textwidth]
{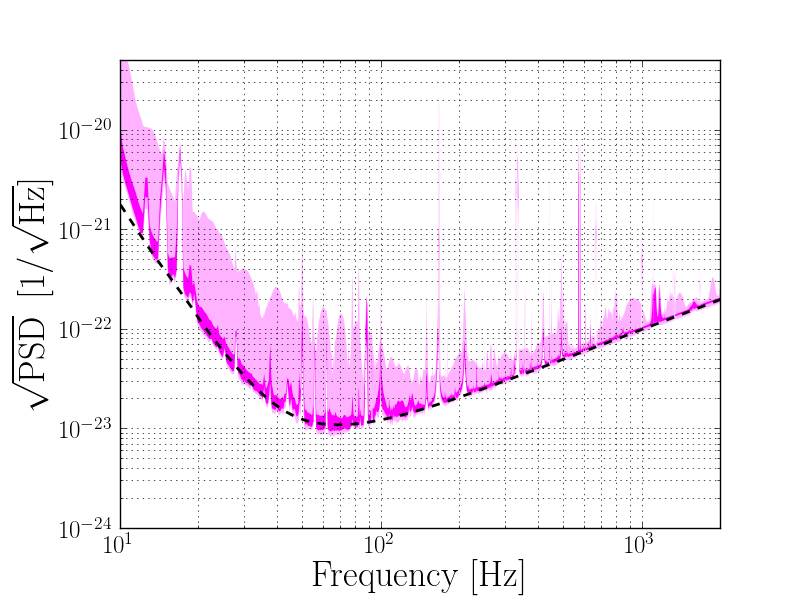}
\caption{\label{fig:NOISE_recolored_sens}
Sensitivity curves of the recolored data for the LIGO Hanford detector (left)
and the Virgo detector (right). In both cases the black dashed line shows the
predicted 2015 -- 2016 sensitivity curve (with the scaling factor added for
Virgo). The dark colored region indicates the range between the 10 \% and 90 \%
quantiles of the PSD over time. The lighter region shows the range
between the minimum and maximum of the PSD over time.}
\end{figure}

In figure \ref{fig:NOISE_recolored_sens} we show some examples of the PSDs
obtained from recoloring the data and compare with
the predicted sensitivity curves. As there are some small stretches of data in
the original science runs where the sensitivity was significantly different
from the average, we show the 10 \% and 90 \% quantiles as well as the maximum 
and
minimum values for the PSD of the recolored data. We notice that the sensitivity
of the detector still varies with time, as in the initial data, and that the
lines in the initial spectra are still present.

\begin{figure}
\centering
\includegraphics[width=0.495\textwidth]
{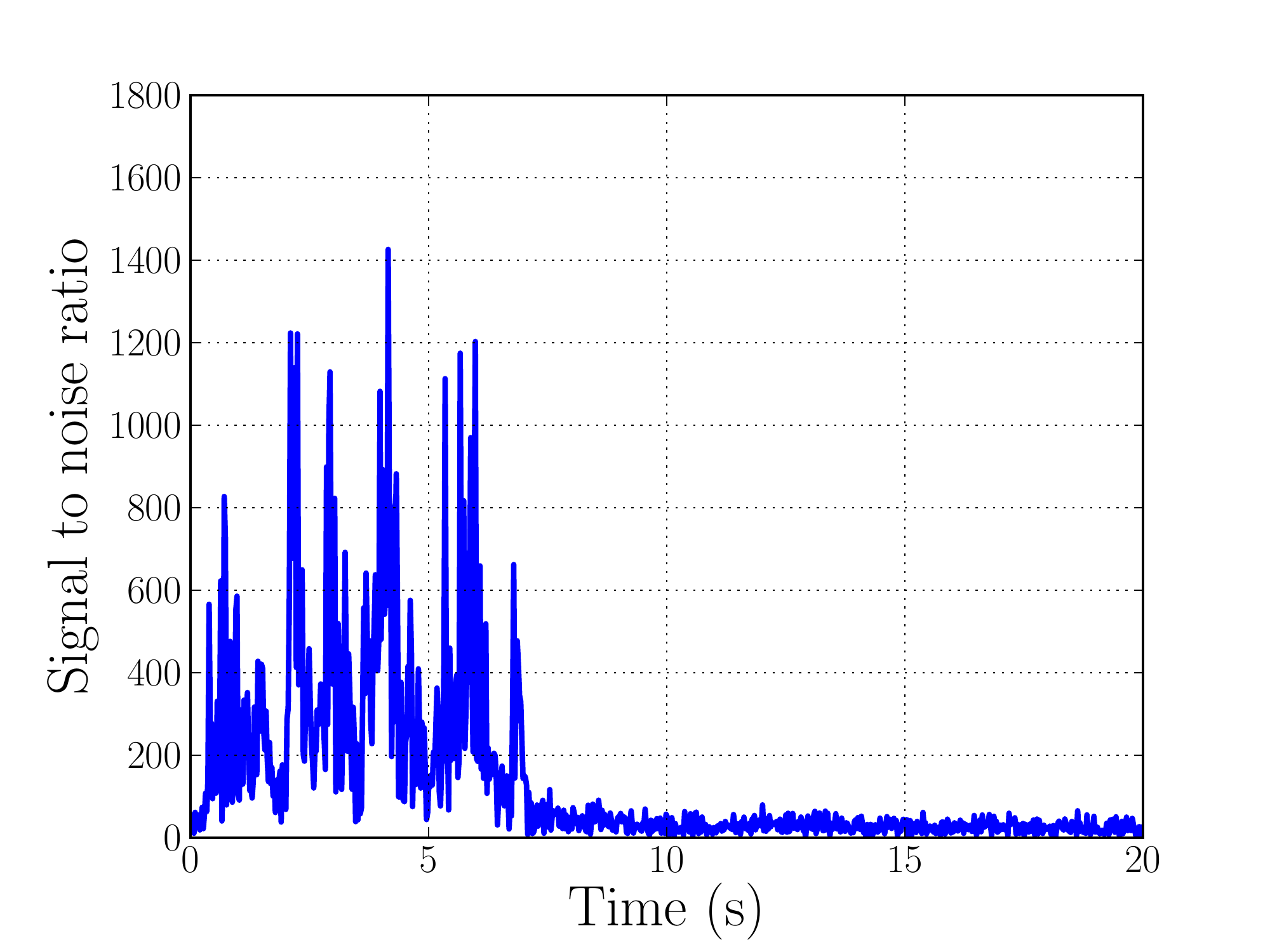}
\includegraphics[width=0.495\textwidth]
{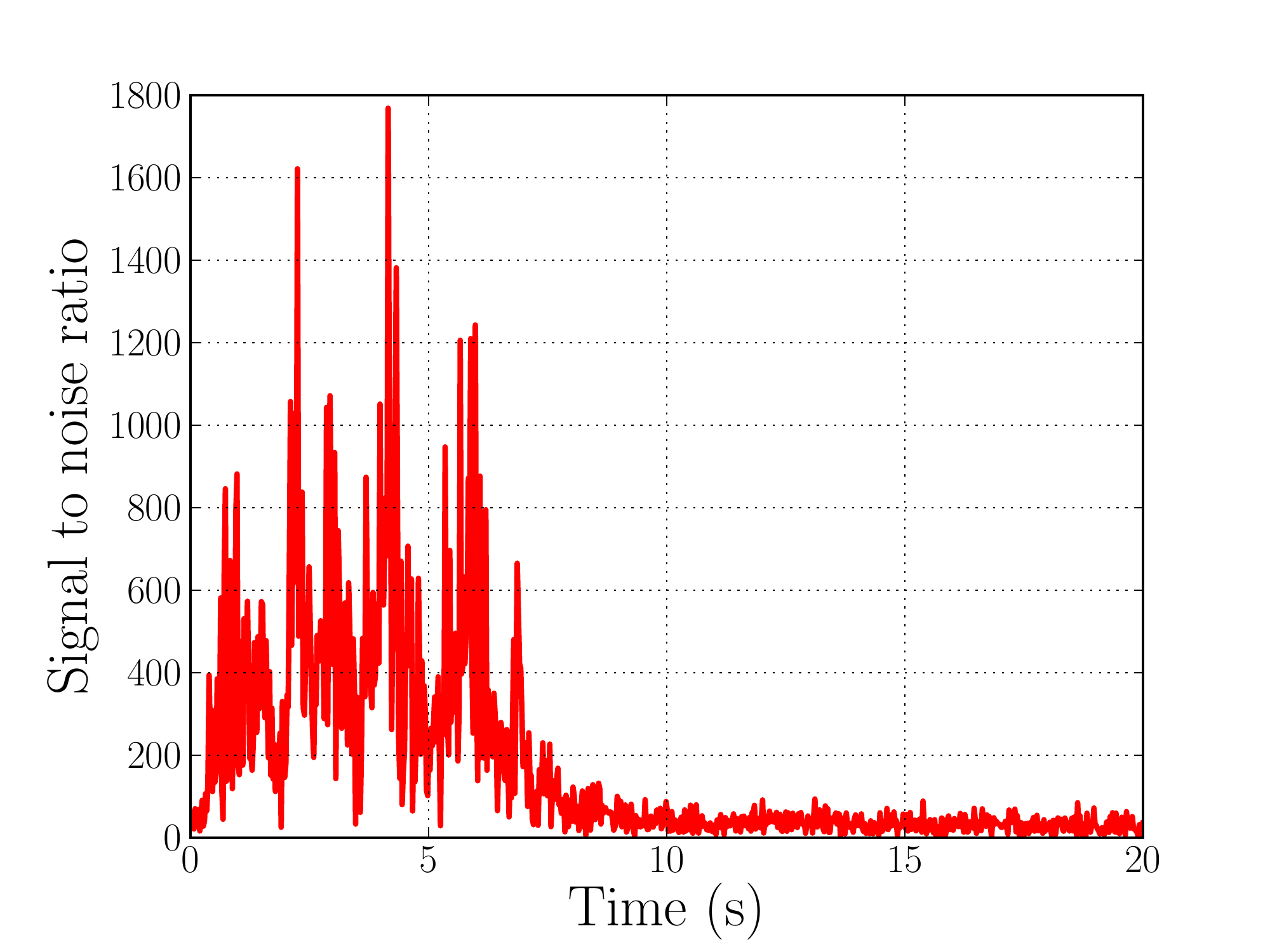}
\caption{\label{fig:NOISE_recolored_glitch}
SNR time series in a 20 s window around a known glitch in the original data
(left) and in the recolored data (right). While the SNR time series clearly
change, the primary features of the glitch are preserved across the recoloring
procedure. These SNR time series were obtained by matched-filtering a short 
stretch of recolored and original data against a (23.7,1.3) $M_{\odot}$ 
template.}
\end{figure}

Non-Gaussian features present in the original data will still be present in
the recolored data, albeit distorted by the recoloring process.
An example of this is shown in figure
\ref{fig:NOISE_recolored_glitch} where we show the SNR time-series around a
known glitch in both the original and recolored data. While the recoloring does
have some effect on the glitch, the two SNR time series are very comparable. As
in searches on the original data, we attempt to mitigate the effect of such
features. A set of \emph{data quality flags} were created for the initial 
detector
data \cite{Aasi:2012wd,LIGOS6Detchar}.
These attempt to flag times where a known instrumental or
environmental factor, which is known to produce non-Gaussian artifacts in the
resulting strain data, was present. To simulate these data quality flags in our
recolored data we simply used the same flags that were present in the original
data and apply them to the recolored data.

\section{Injection Parameters}
\label{sec:parameters}

As an unbiased test of the process through which candidate 
events are identified for BBH waveforms, 7 BBH waveforms were added to the 
recolored data. The analysts were aware that ``blind injections'' had been
added, however the number and parameters of these simulated signals were not 
disclosed until the analyses were completed. This was similar to blind 
injection tests conducted by the LIGO and Virgo collaborations in their latest 
science runs \cite{Colaboration:2011np}.
These injections are self-blinded to ensure that no bias from knowing the
parameters of the signal, or indeed whether a candidate event is a signal or a 
noise artifact, affects the analysis process.

The 7 waveforms added to the data were taken from the numerical
relativity simulations discussed in section \ref{sec:waveforms}. The parameters
of the blind injections are given in Table \ref{tab:blindsigs}. The distribution
of physical parameters used in these blind injections was not intended to
represent any physical distribution. Instead, the injections were chosen to 
test 
the ability
to recover BBH systems across a wide range of parameter space. We describe the 
results of searches for these blind injections in section \ref{sec:searches} 
and 
of parameter estimation studies on these signals in section 
\ref{sec:param_estimation}.

\begin{table}
\caption{\label{tab:blindsigs} The details of the blind injections that were
added to the NINJA-2 datasets prior to analysis. In this table the Event ID
will be used throughout the paper to refer to specific injections. The network
SNR of each injection is denoted by $\rho_{\mathrm N}$. This is the sum of the
overlaps of the injection with itself in each detector, using $30\,$Hz as the
starting frequency in the overlap integrals. $M$ denotes the total mass and $q$
the mass ratio.  $\chi$ denotes the spin on \emph{each} black hole, in all 7
cases both black holes in the binary had the same spin. RA and dec give the
right ascension and declination of the signals respectively.  Dist. denotes the
distance to the source. Detectors online lists the detectors for which data is
present at the time of signal. Hybridization range gives the range of
frequencies in which the signal is hybridized between the post-Newtonian and
numerical components.  Waveform label indicates which numerical waveform was
used, as shown in Tables \ref{tab:ninja2_submissions1} and
\ref{tab:ninja2_submissions2}.  }
\resizebox{\textwidth}{!}{%
\begin{tabular}{@{}ccccccccccc}
\ms\bhline\ms
Event & Waveform   &                    &      &  
  $M$                   &          &  RA     &  Dec.   &  Dist.  &  Detectors & 
Hybrid   \\
ID    & label      & $\rho_{\mathrm N}$ & $q$  &
  ($\mathrm{M}_\odot$)  &  $\chi$  &  (rad)  &  (rad)  &  (Mpc)  &  Online    & 
Range (Hz)        \\
\mr
1 & J4+00+00\_T4  & 23.9 & 4  &  124   &  0.00  &  1.26  &  -0.76  &  569  &  
HLV 
 & 15 -- 18 \\
2 & Ll1-20-20\_T4 & 14.1 & 1  &  35.5  & -0.20  &  1.70  &  -0.03  &  244  &  
HLV 
 & 52 -- 71 \\
3 & Ll1+40+40\_T4 & 16.2 & 1  &  14.4  &  0.40  &  4.18  &  0.07  &  170  &  HLV 
 
& 175 -- 193 \\
4 & G2+20+20\_T4 & 15.1 & 2  &  26.8  &  0.20  &  2.19  &  -0.36  &  247  &  LV  
 
& 68 -- 90 \\
5 & L4+00+00\_T1 & 19.2 & 4  &  19.1  &  0.00  &  1.68  &  0.14 &  83   &  HV  & 
 
86 -- 93 \\
6 & J1+25+25\_T4 & 16.9 & 1  &  75.7  &  0.25  &  4.68  &  0.49 &  854  &  HV  
& 
20 -- 21 \\
7 & J1-75-75\_T1 & 9.8 & 1  &  19.3  & -0.75  &  0.81  &  -0.07 &  292  &  HLV  
& 69 -- 79 \\
\ms\bhline\ms
\end{tabular}}
\end{table}

As well as these blind injections, a large number of (non-blind) simulated 
signals were subsequently analyzed to obtain sufficient statistics to adequately
evaluate the sensitive distances at which the NR waveforms could be detected 
in the early aLIGO and early AdV simulated data sets. For each of the 60 NR 
waveforms given in Table~\ref{tab:ninja2_submissions1} and used in results in 
section \ref{sec:sensitivity}, a set of $\sim42000$ 
simulated signals was generated, necessarily with the same mass ratio and spins 
as the provided NR waveform. The total mass was chosen from a uniform 
distribution between 10 and 100 $M_{\odot}$. The simulations were 
distributed uniformly in distance, however they were not injected beyond a 
distance where they could not possibly be detected. The mass-dependent maximum 
distance that we chose to use is given by
\begin{equation}
 D_{\mathrm{max}} = \left(\frac{\mathcal{M}}{1.219 M_{\odot}}\right)^{5/6}
 \unit[175]{Mpc}.
\end{equation}
Here $1.219 M_{\odot}$ is the chirp mass of a $(1.4+1.4)M_{\odot}$ binary 
system. The factor of $\mathcal{M}^{5/6}$ describes, to leading order, how 
the SNR of the inspiral-only portion of a compact-binary 
merger at fixed distance will scale with mass when the inspiral is 
bandwidth-limited\cite{Peters:1963ux,Allen:2005fk}. $175$ Mpc is chosen because 
it is larger than the distance at which it would be possibly to detect 
a $(1.4+1.4)M_{\odot}$ binary merger with the early noise curves. However, to 
include a large margin for safety $\sim7000$ of the signals were generated with 
chirp-weighted distances between $175$ and $350$ Mpc. The orbital 
orientations, polarization angles and sky directions are all chosen from 
isotropic distributions. The signal coalescence times are drawn from a uniform 
distribution within our analysis window. Coalescence times were limited to 
times where at least two observatories were operating and no data-quality flags 
were active. The results of analyses on the non-blind simulated signals are 
given in section~\ref{sec:sensitivity}.

\section{Search Pipelines}
\label{sec:pipelines}

The goal of this work was to evaluate the detection sensitivity to
binary black hole systems, modelled from the latest numerical
simulations, using the search pipelines that were used to search for
gravitational-wave transient signals in data taken during the final
initial LIGO and Virgo joint observing run.  The two pipelines that
were used to do this were the dedicated compact binary coalescence
(CBC) search pipeline ``\ihope{}'' \cite{Abbott:2009tt, Abbott:2009qj,
Abadie:2011kd, Colaboration:2011np, Aasi:2012rja, Babak:2012zx}
and the unmodelled burst pipeline ``Coherent WaveBurst'' (cWB)
\cite{Abbott:2007wu, Abadie:2010mt, Abadie:2012rq, Virgo:2012aa}. The
\ihope{} pipeline was developed as a search pipeline for detecting
compact binary mergers. It employs a matched-filtering algorithm
against a bank of template waveforms~\cite{Babak:2012zx}. The \ihope{}
pipeline was used to search for CBC systems (not just binary black
holes) with component masses $\in [1,99]\, \mathrm{M}_{\odot}$.  As a
complement to template-based specialized searches, cWB was developed
as an all-purpose un-modeled search pipeline, hence, it does not
require \emph{a priori} knowledge of the signal waveforms. It is
better suited for burst signals spanning a small time-frequency
volume. Moreover, due to the lack of model constraints, cWB is more
adversely affected by background noise than matched-filter searches.
Past simulation studies with initial LIGO sensitivity curves have
shown that cWB was sensitive to CBC mergers with total masses $\in
[25,500]\, \mathrm{M}_{\odot}$ over wide regions of the binary
parameter space \cite{Pankow:2009nx}.

In addition to the \ihope{} and cWB detection pipelines we also use parameter 
estimation algorithms to provide estimates of the parameters of compact binary 
systems observed with the detection algorithms. 
In the following section we provide a brief overview of the detection and 
parameter estimation pipelines.
The results of running these searches on the data containing the NINJA-2 blind
injections are presented in section \ref{sec:searches} and parameter estimation 
results given in section \ref{sec:param_estimation}.

\subsection{Coherent WaveBurst}
\label{ssec:cwb_pipelines}

Coherent WaveBurst is a multi-resolution algorithm for coherent
detection and reconstruction of gravitational wave
bursts~\cite{Klimenko:2008fu}. The cWB algorithm has been used in
various LIGO-Virgo burst searches
\cite{Abbott:2007wu,Abadie:2010mt,Abadie:2012rq} and more recently in
the search for intermediate mass black hole binaries
\cite{Virgo:2012aa}.  Within the framework of the constrained maximum
likelihood analysis \cite{Klimenko:2008fu}, cWB identifies GW signals
in data from multiple detectors and provides estimates of the signal
parameters, e.g. sky location and waveforms.  Along with the
reconstruction of un-modeled burst signals, which imply random
polarization, cWB can perform loosely modeled likelihood analyses
assuming different polarization states, i.e. elliptical, linear or
circular.

The NINJA2 cWB analysis uses the elliptical polarization constraint 
\cite{Pankow:2009nx,Virgo:2012aa} and searches for signals in the frequency 
band from 32 Hz to 1024 Hz. The analysis is performed in several steps: first, 
the data streams from all GW detectors are processed with the Meyer's wavelet 
transformations with 6 different  time-frequency resolutions of 
$ 4\times1/8$, $8\times1/16$, $16\times1/32$, $32\times1/64$, 
$64\times1/128$, $128\times1/256$ [Hz $\times$ s]. Then the data are 
conditioned 
with a linear predictor error filter to remove power lines, violin modes 
and other predictable data components. Triggers are reconstructed as the 
coherent
sets of samples (pixels) identified in the time-frequency data. For each 
trigger the coherent 
statistics are then computed. These include the network correlation 
coefficient, ${cc}$ and the network energy disbalance, ${\Lambda}$, which are 
used to enable the signal consistency selection cuts. The cWB detection 
statistic is the coherent network amplitude, $\eta$, which is used to rank the 
events and thereby establish the significance against a sample of 
background events obtained with the time-shift analysis 
\cite{Klimenko:2008fu,Virgo:2012aa,Pankow:2009nx}. 
This shifting procedure is typically performed 
thousands of times in order to accumulate sufficient statistics.

\subsection{\ihope{}}
\label{ssec:ihope_pipelines}

The \ihope{} pipeline is designed to search for gravitational waves
emitted by coalescing compact binaries~\cite{Babak:2012zx}. It has been
optimized for and used in LIGO and Virgo GW searches over the past 
decade~\cite{Abbott:2007xi, Abbott:2009tt, Abbott:2009qj, Abadie:2010yb, 
Colaboration:2011np, Aasi:2012rja}, and also in the mock Laser 
Interferometer Space Antenna (LISA) data 
challenges~\cite{Babak:2008aa}. The NINJA-2 \ihope{} analysis uses the same 
pipeline-tuning that was used in the searches performed during 
the final initial LIGO and Virgo joint observing 
run~\cite{Colaboration:2011np}.

The pipeline matched-filters the detector data against a bank of
analytically modelled compact binary merger 
waveforms~\cite{Allen:2005fk,Babak:2012zx}. Only nonspinning compact binary 
merger signals are used as filters and the bank is created so as to densely 
sample the range of possible binary masses~\cite{Babak:2006ty}.
For each detector, the filtering stage produces 
a sequence of \textit{triggers} which are plausible events with a high 
signal to noise ratio SNR $\rho$. The algorithm proposed 
in~\cite{Robinson:2008un} is used to keep only those that are found coincident
in more than one detector across the network, which helps remove triggers due to 
noise. 
Knowledge of the instrument and its environment is used to further exclude
triggers that are likely due to non-Gaussian noise transients, or 
\textit{glitches}. Periods of heightened glitch rate are
removed (\emph{vetoed}) from the analysis. The time periods where the rate of 
glitches is elevated are divided into $3$ \emph{veto categories}. Periods of 
time flagged by category 1 and 2 vetoes are
not included in the analysis as known couplings exist between instrumental
problems and the gravitational-wave channel during these periods. Periods of
time vetoed at category 3 are \emph{likely} to have instrumental problems. A
strong gravitational-wave signal can still be detected during category 3 times,
but including these periods in the background estimate can compromise our
ability to detect weaker signals in less glitchy periods of time. For this
reason the search is performed both before and after category 3 vetoes are 
applied. The significance of events
that survived category 1-3 vetoes were calculated using the background that also
survived categories 1-3. The significance of events that survived category 2
but were vetoed at category 3 were calculated using background that survived
categories 1-2.

Signal based 
consistency measures further help distinguish real signals from background noise
triggers in those that are not vetoed and pass the coincidence test. 
The $\chi^{2}$ statistic proposed in~\cite{Allen:2004gu} quantifies the 
disagreement in the frequency evolution of 
the trigger and the waveform template that accumulated the highest SNR
for it, c.f. Eq.~(4.14) of~\cite{Allen:2004gu}. We weight the SNR with this
statistic to obtain the \textit{reweighted} SNRs for all coincident triggers. 
The exact weighting depends on the mass range the search is focused on, c.f.
Eq.~(17,18) of~\cite{Babak:2012zx}. The reweighted SNR is used as the
ranking statistic to evaluate the significance, and thus the false alarm rate 
(FAR), of all triggers. 

Following the division of the mass-parameter space used 
in~\cite{Colaboration:2011np, Aasi:2012rja}, we performed both \emph{low 
mass} and \emph{high mass} \ihope{} searches on the NINJA-2 data. The low-mass 
search focused on binaries with $2M_\odot
\leq m_{1}+m_{2} < 25M_\odot$, and used frequency domain $3.5$PN 
waveforms as templates~\cite{Blanchet:2001ax, Blanchet:2001aw, Blanchet:2004ek}.
The high-mass search instead focused on the mass-range 
$25M_\odot \leq m_{1}+m_{2} < 100M_\odot$, and used the effective-one-body
inspiral-merger-ringdown model calibrated to numerical relativity, as described 
in~\cite{Buonanno:2007pf}. The exact $\chi^2$-weighting used to define the
re-weighted SNR varied between the two analyses~\cite{Babak:2012zx}. The 
significance of the triggers found by both was estimated as follows.
All coincident triggers are divided into $4$ 
categories, i.e. HL, LV, HV and HLV, based on the detector combination they are 
found to be coincident in~\cite{Colaboration:2011np}. They are further divided 
into
$3$ \textit{mass}-categories based on their chirp mass
$\mathcal{M}_{c}=(m_{1}m_{2})^{3/5}(m_{1}+m_{2})^{-1/5}$ for the low-mass 
search, and $2$ categories based on their length in time for the high-mass 
search~\cite{Colaboration:2011np}. The rate of background noise triggers, or 
\textit{false alarms}, has been found to be significantly higher for shorter 
signals from more massive binaries, and also to be different depending on the 
detector
combination, and these categorizations help segregate these effects for 
estimation of
the background~\cite{Colaboration:2011np,Babak:2012zx}. For all the triggers 
the combined
re-weighted SNR $\hat{\rho}$ is computed, which is the quadrature sum of 
re-weighted SNRs across the network of detectors. All triggers are then 
ranked in each of the mass and duration sub-categories 
independently according to their $\hat{\rho}$, allowing us to put a limit on the 
trigger
false alarm rate (FAR) at a given threshold $\hat{\rho}=\hat{\rho}_{0}$. This 
is 
described by 
\begin{equation}\label{eq:FARdef}
\mathrm{FAR}~(\hat{\rho}_{0}) \leq \frac{N(\hat{\rho}\geq 
\hat{\rho}_{0})+1}{T_{c}},
\end{equation}
where $N(\hat{\rho}\geq x)$ is the number of background noise triggers with 
$\hat{\rho}$
greater than or equal to $x$, and $T_{c}$ is the total time analyzed 
for that coincidence category. From 
~\ref{eq:FARdef}, the smallest FAR we can estimate is $1/T_{c}$, and to get a 
more precise
estimate for our detection candidates we simulate additional background time.
We shift the time-stamps on the time-series of single detector triggers by
$\Delta t$ relative to the other detector(s),
and treat the shifted time-series as independent coincident background time. All 
coincident
triggers found in the shifted times would be purely due to background noise. 
We repeat this process setting $\Delta t=\pm 5s, \pm 10s, \pm 15s,\dots$, 
recording all the
time-shifted coincidences, until $\Delta t$ is larger than the duration of the 
dataset itself. 
With the additional coincident background time $T_{c}$ accumulated in this way, 
we can get a more precise estimate of the low FARs we expect for detection 
candidates, which are described in detail in section~\ref{ssec:ihope_results}.

\subsection{Parameter estimation}
\label{ssec:PE_pipelines}

The detection methods described above produce times of interest where a
gravitational wave may be present in the data (i.e. triggers), along with point
estimates of the compact object masses from the signal, independently in each 
detector. These triggers are followed up with the goal of
estimating the posterior probability density function of the parameters
that describe the signal and to evaluate the evidence of different
waveform models. In order to do so, we use Bayesian methods, in which
the data from all detectors are analysed coherently.

The Bayesian parameter estimation
algorithms used in this work provide estimates of the posterior probability 
distribution function.
The probability of a set of parameters $\vec{\theta}$ under a model $M$ given
the observed data $d$ can be written as
\begin{equation}
    \label{posterior}
    p(\vec{\theta}|d,M) = \frac{p(\vec{\theta}|M)p(d|\vec{\theta},M)}{p(d|M)}.
\end{equation}
Here $p(d|\vec{\theta},M)$ is the likelihood of observing the measured data
given the set of parameters $\vec{\theta}$, and $p(d|M)$ is the marginal
distribution of the data under model $M$, commonly referred to as the evidence.
When only concerned with parameter estimation the evidence is a normalization
constant that can be ignored.  It becomes relevant however, when comparing how
well two models do in describing the observed data.  By marginalizing over all
model parameters the evidence is obtained
\begin{equation}
    \label{evidence}
    p(d|M) = \int p(\vec{\theta}|M)p(d|\vec{\theta},M) d\vec{\theta}.
\end{equation}
Given the evidence of two competing models, $M_1$ and $M_2$, the support for
$M_1$ over $M_2$ can be quantified via the Bayes factor
\begin{equation}
    \label{bayesFactor}
    \mathcal{B}_{12} = \frac{p(d|M_1)}{p(d|M_2)}.
\end{equation}

These techniques require the generation of $\sim 10^6 - 10^7$ model waveforms
to probe the 9 (for non-spinning black holes) to 15 (for fully spinning black
holes) dimensional parameter space of compact binary systems in circular orbit,
making it infeasible at present to use numerical relativity simulations.
Instead approximate models (\textit{e.g.} post-Newtonian, effective one-body)
that are computationally cheaper to produce are used to estimate the parameters
of a measured signal. Numerous studies have assessed the statistical
uncertainty in compact binary parameter
estimates~\cite{vanderSluys:2009bf,Raymond:2008im,Veitch:2009hd,   
Raymond:2009cv}, which use the same approximate model for injection and
analysis. Few studies have been done to quantify the systematic uncertainty in
parameter estimates due to the use of these approximate models
\cite{Canitrot:2001hc,Cutler:2007mi}. Numerical relativity simulations
provide us with the most accurate waveforms currently available, making them
ideal for quantifying the systematic uncertainties inherent with using
approximate models.  This mock data challenge is the first time such a study
has been conducted using models that account for the component angular momenta
of the compact objects.

The Markov Chain Monte Carlo sampler
\texttt{lalinference\_mcmc}~\cite{vanderSluys:2007st,vanderSluys:2008qx}, and
two nested sampling implementations
\texttt{lalinference\_nest}~\cite{Veitch:2009hd} and
\texttt{MultiNest}~\cite{Feroz:2008xx} from the \textit{LALInference} package
of the LSC Algorithm Library~\cite{LAL} were used to follow up GW candidates
from the detection search pipelines. Due to the computational burden, we carried
out the analysis with \texttt{lalinference\_mcmc}, and as a consistency check
for selected candidates and waveforms posterior estimates were also obtained
with \texttt{lalinference\_nest} and \texttt{MultiNest}. For model comparisons
we have calculated the evidence by marginalizing each posterior estimate using
thermodynamic integration.

Each candidate was analyzed using two distinct waveform models: \imr and
\eob.  Both models describe the IMR phases of the GW from a compact binary 
merger.  \eob models
non-spinning binaries using the effective-one-body (EOB) 
that re-sums the PN dynamics and energy flux, and describes 
the merger-ringdown signal as a superposition of 
quasi-normal modes~\cite{Pan:2011gk}. \imr is a
phenomenological model with a PN description of the inspiral phase
building up on test-mass terms to 2PN order, fit to a set of spinning and 
non-spinning PN-NR hybrid waveforms~\cite{Ajith:2009bn}.  Waveforms are 
generated in the frequency
domain and model binaries with component spins aligned with the orbital
angular momentum through a single spin parameter $\chi \equiv
(1+\delta)\chi_1/2 + (1-\delta)\chi_2/2$. Here $\delta \equiv (m_1-m_2)/M$ and
$\chi_i\equiv S_i/m_i^2$, where $S_i$ denotes the angular momentum of the $i$th
component of the binary, and $M$ the total mass of the system. 

The mass-ratio dependent and higher order terms used in \imr are calibrated to 
PN-NR hybrids
that cover the late-inspiral, merger and ringdown. Therefore the accuracy of the
model is expected to decline with decreasing total mass, as the inspiral phase
of the waveform becomes a larger fraction of the total SNR of the signal,
especially for comparable mass binaries. At the time of the analysis however, 
it was the only IMR waveform, including spin effcts, that was computationally
feasible for use, making it the most physically relevant waveform for the 
analysis. \eob is more computationally expensive, but has been
shown to be accurate enough for uncertainties in parameter estimates to be
dominated by statistical error, rather than
systematic~\cite{Littenberg:2012uj}.  It only models binaries with non-spinning
components however, so the model is only relevant for non-spinning injections.
\texttt{SEOBNRv1} is the successor to \eob that accounts for
(aligned) spin~\cite{Taracchini:2012ig}, however it is currently too 
computationally expensive to be used for parameter estimation.

Due to the lack of astrophysical constraints on compact binary systems, it is
difficult to physically motivate any particular choice for the prior 
distribution of the intrinsic parameters (i.e. masses and spins).  For this 
study we have
chosen to use distributions that are uniform in component masses and component 
spin magnitudes over the range of parameter values being injected.  The prior
distribution was also flat in coalesence time across a 200 ms window centered
on the trigger time, isotropic in orientation angles (e.g. inclination), and
volumetric, giving equal prior probability to all spatial locations.

\section{Blind Injection Challenge Results}
\label{sec:searches}

In this section we present the results of using the detection pipelines 
described in section \ref{sec:pipelines} to search for the blind injections 
listed 
in Table \ref{tab:blindsigs}.

\subsection{Coherent WaveBurst}

For the NINJA2 cWB analysis 
it was decided \textit{a priori}
to search for GW bursts in the entire available 
times during which all three detectors were operating
(17.9 days) and to discard the remaining times.
First the search was performed on a total of 
12,000 time-lagged observation times, accumulating 563.7 years of effective 
background live time. The background events that survived the data quality and 
analysis selection cuts (i.e. ${cc}>0.7$ and ${\Lambda}<0.4$) 
were used for calculation of the significance of candidate events. This 
background sample contains all reconstructed events, most of which do 
not resemble expected compact coalescence waveforms, as we do not 
enforce any waveform model. As a result, our background distribution is 
populated by relatively high signal-to-noise ratio events.   

After completing the background analysis, the zero lag live 
time was analysed. The search detected an on-source event showing a chirping 
waveform compatible with a compact binary coalescence at a SNR 
$\sim 22.1$ and $\eta = 7.1$ . The FAR of the candidate was 
estimated at $\sim 1/47~yr^{-1}$ from
comparison with the burst reference background, yielding a false alarm 
probability (FAP) of $\sim$ 0.001. After the parameters of the blind injections
were disclosed, this event was revealed to be the first blind injection of 
Table 
\ref{tab:blindsigs}.
As a follow-up analysis, we investigated \textit{a posteriori} all the times of 
the blind injections, as well as those on 2-fold exclusive live time. We found 
that the rest of the injected signals are either reconstructed with extremely 
low $\eta$ or missed, see Table \ref{tab:cWBEvents}. For massive systems, such 
as events 1 and 6, the cWB algorithm recovers a large fraction of the 
injected signal-to-noise ratio. For lighter binaries, as expected, the 
algorithm is largely sub-optimal
\footnote{Lately, a lot of work has been 
devoted to extend the sensitivity of the algorithm to lower total masses, which 
is part of the on-going upgrades of cWB.}.     

\begin{table}
\caption{\label{tab:cWBEvents} The cWB search and follow-up results.
The Event Labels correspond to those of each blind injection given in
Table \ref{tab:blindsigs}. This association is based on the time of
the candidates relative to the time of the injections. $M$ denotes the
total mass. The false alarm probability, FAR, and false alarm probability, FAP, 
of each event are
estimated by comparison with the empirically-calculated background
distribution of the corresponding network of detectors. All but the first event 
are well within the bulk of the
corresponding FAR distributions. $\eta$ is the network correlated amplitude, 
which is
the main cWB detection statistic.  The injected network SNR
($\rho_{inj}^{net}$) is the square root of the quadratic sum of the
optimal SNR in each detector. The recovered network SNR
($\rho_{rec}^{net}$) is the cWB estimate of the injected network SNR.
Events 5 and 7 were completely missed due to a low reconstructed SNR
and/or because of nearby noise glitches.}
\begin{indented}
\item[]\begin{tabular}{@{}ccccccccc}
\ms\bhline\ms
 \subrows{Event\\ID} &    \subrows{Event\\Label} &  $M$ & \subrows{1/FAR\\(yr)} 
& FAP & Network & 
$\eta$ & $\rho_{rec}^{net}$ & $\rho_{inj}^{net}$ \\
 \mr
 1 &  J4+00+00\_T4 & 124 & 47 & 0.001 & HLV & 7.1 & 22.1 & 22.8 \\ 
 2 &  Ll1-20-20\_T4 & 35.5 & -  & -   & HLV & 2.8 & 9.1  & 13.9 \\  	
 3 &  Ll1+40+40\_T4 &  14.4 & -  & -   & HLV & 2.7 & 9.2  & 15.7 \\
 4 &  G2+20+20\_T4 & 26.8 & -  & -   & LV  & 1.6 & 7.4  & 14.1 \\
 5 &  L4+00+00\_T1 & 19.1 & -  & -   & HV  & - & - & 18.5 \\
 6 &  J1+25+25\_T4 & 76.7 & -  & -   & HV  & 2.0 & 13.8 & 15.9 \\
 7 &  J1-75-75\_T1 & 19.3 & -  & -   & HLV  & - & - & 9.5 \\
\ms\bhline\ms
\end{tabular}
\end{indented}
\end{table}

\subsection{\ihope{}}
\label{ssec:ihope_results}

\begin{table}
\caption{\label{tab:ihopeEvents}The \ihope{} search results. The Event IDs
correspond to the Event ID of each blind injection given in Table
\ref{tab:blindsigs}; this association is based on the time of the candidates
relative to the time of the injections. The FARs are
calculated from all possible $5\,$s time shifts in a two-week period
surrounding each event. $M$ and $q$ give the total mass and mass ratio 
respectively that were recovered in each detector. The
recovered SNR ($\rho_{\mathrm{rec}}$) and re-weighted SNR ($\hat{\rho}$) are
reported separately for each detector. To calculate FARs the quadrature sum of
$\hat{\rho}$ was used. Unless noted, FARs were calculated after category 1-3
vetoes were applied.}
\begin{indented}
\item[]\begin{tabular}{@{}ccccccccc}
\ms\bhline\ms
    \subrows{Event\\ID} & \subrows{1/FAR\\(yr)} & Detectors & $M$ & 
$q$ & $\rho_{\mathrm{rec}}$ & $\hat{\rho}$ & 
Search \\
\mr
    1 & $\geq 6200$ & \subrows{H\\L} & \subrows{99.8\\94.7} & 
\subrows{24.7\\13.8} & \subrows{18.6\\14.5} & 
\subrows{12.3\\13.2} & High mass \\ 
\mr
    2 & $\geq 10000^*$ &  \subrows{H\\L\\V} & \subrows{37.1\\42.7\\40.3} & 
\subrows{3.22\\3.99\\2.47} &
\subrows{6.6\\9.6\\9.2} 
& \subrows{4.9\\9.6\\9.2} & High mass \\ 
\mr
    \multirow{2}{*}{3} & $\geq 23000$ & \subrows{L\\V} & \subrows{13.8\\14.2} & 
\subrows{1.15\\1.41} & \subrows{12.4\\5.9} & 
\subrows{11.6\\5.2} & \subrows{Low mass\\Cat. 3}\\ \cline{2-8}
      & $\geq 5800^{\dagger}$ &  \subrows{H\\L\\V} & \subrows{13.7\\13.8\\14.2} 
& \subrows{1.15\\1.15\\1.41} & \subrows{7.9\\12.4\\5.9} & 
\subrows{7.5\\11.6\\5.2} & \subrows{Low mass\\Cat. 2} 
\\ 
\mr
    \multirow{2}{*}{4} & $\geq 31000$ & \subrows{L\\V} & \subrows{24.8\\25.5} & 
\subrows{1.14\\1.55} &
\subrows{9.0\\12.4} & \subrows{8.7\\12.4} & High mass \\ \cline{2-8}
      & $\geq 23000$ & \subrows{L\\V} & \subrows{25.0\\25.0} & 
\subrows{1.80\\1.43} & \subrows{8.5\\10.9} & 
\subrows{8.5\\9.2} & Low mass \\ 
\mr
    5 & $\geq 21000$ & \subrows{H\\V} & \subrows{19.5\\22.2} & 
\subrows{4.27\\6.24} & \subrows{16.2\\8.8} & 
\subrows{15.6\\8.1} & Low mass \\ 
\mr
    6 & $\geq 37000$ & \subrows{H\\V} & \subrows{72.4\\65.8} & 
\subrows{5.19\\1.19} & \subrows{10.6\\14.7} & 
\subrows{10.6\\12.0} & \subrows{High mass\\Cat. 2} \\
\mr
    7 & \multicolumn{7}{c}{\it{Not found}}  \\ 
\ms \bhline\ms
\end{tabular}
\\
$^*$ Only used LV triggers for computing significance of this
event; see section \ref{ssec:ihope_results}. \\
$^\dagger$ Only used HL triggers for computing significance of
this event; see section \ref{ssec:ihope_results}.
\end{indented}

\end{table}

The results of the low-mass and high-mass \ihope{} searches are presented in
Table \ref{tab:ihopeEvents}. The Event IDs correspond to the Event IDs of the
blind injections in Table \ref{tab:blindsigs}. The mapping between the \ihope{}
candidates and the blind injections is based on the event times of each. All
injections except for injection 7 were found with high significance in one or
both searches. Event 7 was missed because the injection's SNR was too small to
be detected by the pipeline. The optimal SNR of this injection --- obtained
by finding the overlap of the injection with itself --- was $5.7$ in H, $6.0$ in 
L,
and $5.3$ in V, giving a network SNR of $9.8$. However, the injected SNR in 
Virgo was
below the SNR threshold used by the \ihope{} pipeline (= $5.5$). This
means that, at best, the event could only surpass threshold in H and L, giving
a maximum recoverable network SNR of $8.2$; the false alarm rate at this
network SNR is order $10^3$ per year.

For this analysis we used the same vetoes as were used in
\cite{Colaboration:2011np} and \cite{Aasi:2012rja} applied to
the corresponding times in the recolored data. After veto categories
1-3 were applied, the total analyzed time consisted of $0.6$ days of
coincident HL data, $5.4$ days of coincident LV data, $6.5$ days of
coincident HV data and $8.9$ days of coincident HLV data.  FARs were
calculated in each bin using the time-shift method described in
section \ref{ssec:ihope_pipelines}, then combined over all bins.

Table \ref{tab:ihopeEvents} also gives the total masses and mass ratios that 
were recovered by the \ihope{} pipeline in each detector for each candidates. 
We see that the values reported by \ihope{} can vary substantially
from the injected parameters. This is not surprising: many of the injections
had spin, and one injection (Event 1) was outside of the mass range covered by 
the template bank.
We also see that the high-mass search deviates from the actual mass parameters
more than the low-mass search. This, too, is expected since the template bank in
the high-mass search is more sparsely populated. In general, templates are
placed in \ihope{} so as to maximize detection probability across the parameter
space while minimizing computational cost. \ihope{} therefore only provides a
rough estimate of candidate parameters. For more precise estimates we use the
parameter estimation techniques described in section \ref{ssec:PE_pipelines}, 
the
results of which are presented in section \ref{sec:param_estimation}. For 
compact binary systems where most of the SNR is obtained from the inspiral, 
\ihope{} is expected to give a good estimation of the chirp mass of the 
system~\cite{Cutler:1994ys,Poisson:1995ef,Hannam:2013uu}. For the lowest mass 
systems in this study, \ihope{} is not recovering the chirp mass to within 5\% 
accuracy. For the higher mass systems the error on the chirp mass recovered by 
\ihope{} grows to over 100\% for event 1.

\begin{figure}
  \centerline{\includegraphics[width=\columnwidth]{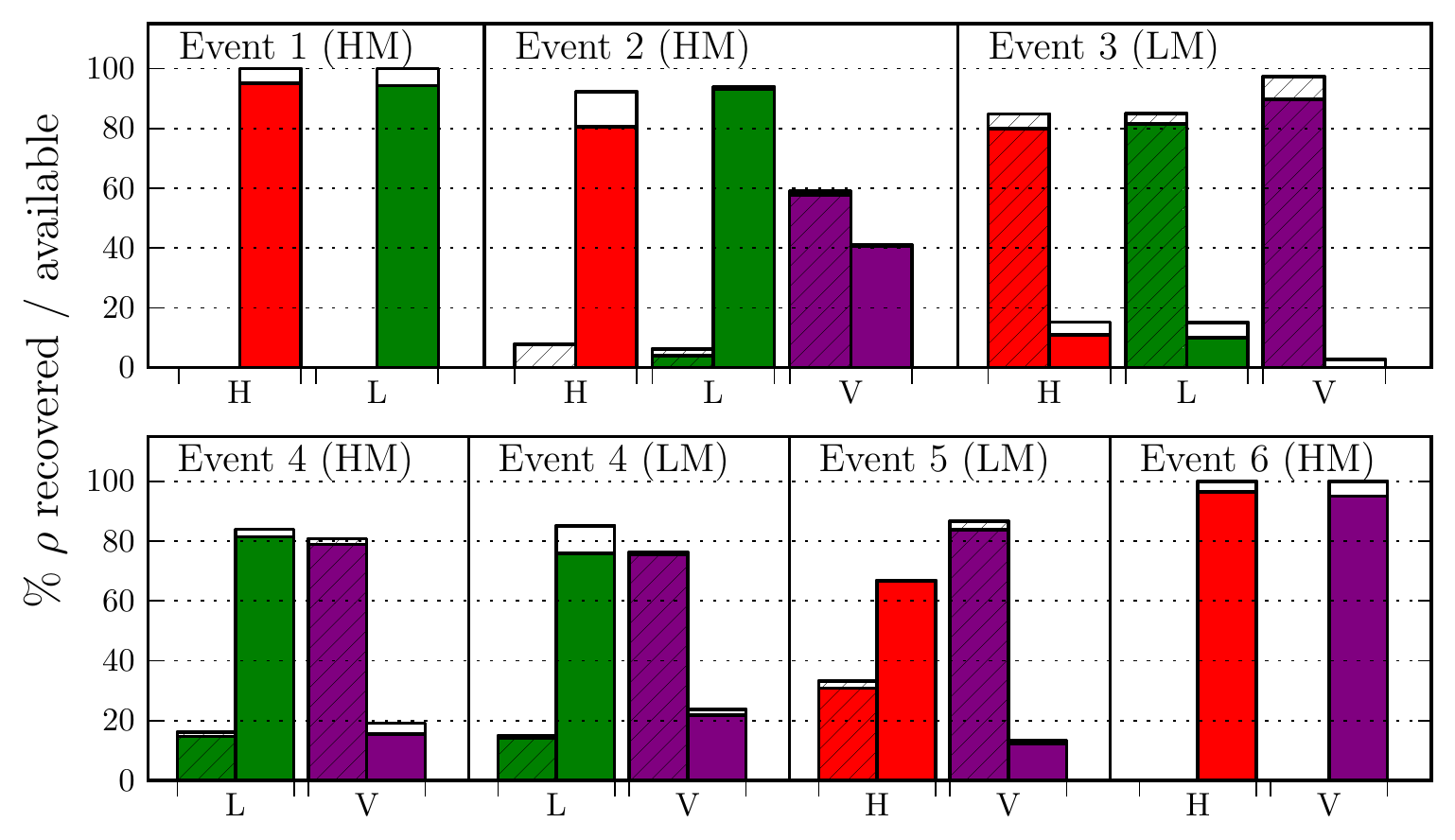}}
  \caption[SNR distribution]{\label{fig:snrDistribution}
SNR recovered by \ihope{} as a fraction of total available SNR in each detector
for each found injection. Hatched bars give the percentage of available SNR in
the PN part of the injection; solid bars give the percentage of available SNR
in the NR part. Color bars indicate the amount of SNR recovered by \ihope{}.
``LM" indicates events recovered by the low-mass search --- which used 3.5PN
waveforms for templates --- ``HM" indicates events recovered by the high-mass
search --- which used EOBNRv1 waveforms for templates. The PN SNR is determined
by terminating the matched filter at the frequency half-way between the
hybridization range; the NR part is given by filtering from that frequency and
up.  Remarkably, both the low-mass and high-mass searches recovered most of the
NR SNR despite the templates not having merger and ringdown (low mass) or not
calibrated to the numerical waveforms used for the injections (high mass).
  }
\end{figure}

The greatest concern for a detection pipeline like \ihope{} is whether 
the mismatch between templates and signals is small enough so as not to lose a 
substantial amount of re-weighted SNR. The templates used in this 
search were able to recover enough SNR of the blind injections to make them 
stand
significantly above background. One might think that the majority of the
recovered SNR comes from templates matching the PN part of the injected
waveforms.  However, figure \ref{fig:snrDistribution} shows the SNR recovered by
\ihope{} in the PN and NR parts of the injections as a fraction of the total
available SNR.  We see that most of the available NR SNR is recovered in every
event even though template waveforms did not have merger and ringdown (in the
case of events recovered by the low-mass search), or were not calibrated to
these particular numerical waveforms (in the case of events recovered by the
high-mass search). To more rigorously determine what effect mismatch between
templates and signals may have on detection sensitivity, we perform a large
scale injection campaign with the NINJA-2 waveforms in section 
\ref{sec:sensitivity}.

Initially we used 100 time shifts to identify candidate events. All of the
coincident events associated with the blind injections were louder than
all background in the 100 time shifts. These were the only events to be
louder than all background. Using 100 time shifts we could only bound the FAR
of the events to $\lesssim 10\,yr^{-1}$, which is not small enough to claim a
detection. To improve our estimate, we performed as many $5\,$s time shifts as
possible in the NINJA-2 dataset. This is the same
method that was used for the blind injection described in
\cite{Colaboration:2011np}. 

Two blind injections were found in all three detectors: Event $2$ in the
high-mass search and Event $3$ in the low-mass search (before category 3 vetoes
were applied).  Estimating background using the extended slide method with
three detectors adds computational complexity, and has not previously been
performed (the blind injection in \cite{Colaboration:2011np} was only
coincident in two detectors). However, in both Events 2 and 3 one of the three
detectors had significantly less $\hat{\rho}$ than the other two (H in Event 2
and V in Event 3). We therefore did not include the detector with the smallest
$\hat{\rho}$ when estimating the extended background for these two events. 

Event 6 was vetoed at category 3. We therefore calculated its significance only
after the first two veto categories were applied. All of the other events
survived category 3 vetoes. Na\"{i}vely, we expect these events to have lower
FARs if their significance is calculated after categories 1-3 have been
applied. However, for Event 3 the trigger in the H detector was vetoed at
category 3, leaving only L and V. Since the H trigger contributed a substantial
amount of the combined re-weighted SNR, we might expect the resulting FAR to be 
\emph{higher} for this event after category 3. A method to deal 
with partially vetoed events like this has not been proposed. We therefore 
simply report both results here.

Event 4 was found with high significance by both the high-mass and low-mass
searches. This is not surprising as the injected total mass was
$26.8\,\mathrm{M}_\odot$, which is close to the boundary between the two
searches. Currently no method has been established on how to combine the
results from the low-mass and high-mass searches. We therefore give both
results here.

\section{Parameter Estimation Results}
\label{sec:param_estimation}

For each of the blind injections described in section~\ref{sec:parameters} and 
\ref{sec:searches}, parameter distributions were estimated using both 
non-spinning and spin-aligned models, as functions of eleven and nine 
parameters, respectively.  \imr was used both
as an aligned-spin model, as well as a non-spinning model by fixing the
effective spin $\chi$ to 0, which we will refer to as \imrns.  In addition, the
\eob model was used as an additional non-spinning waveform (see
section \ref{ssec:PE_pipelines} for model descriptions).  From these posterior
estimates, one can determine the marginalized distributions in
parameters of interest, as well as the evidence for the given model
(see eq.~(\ref{evidence})).

\begin{figure}[t] 
  \includegraphics[width=0.9\textwidth]{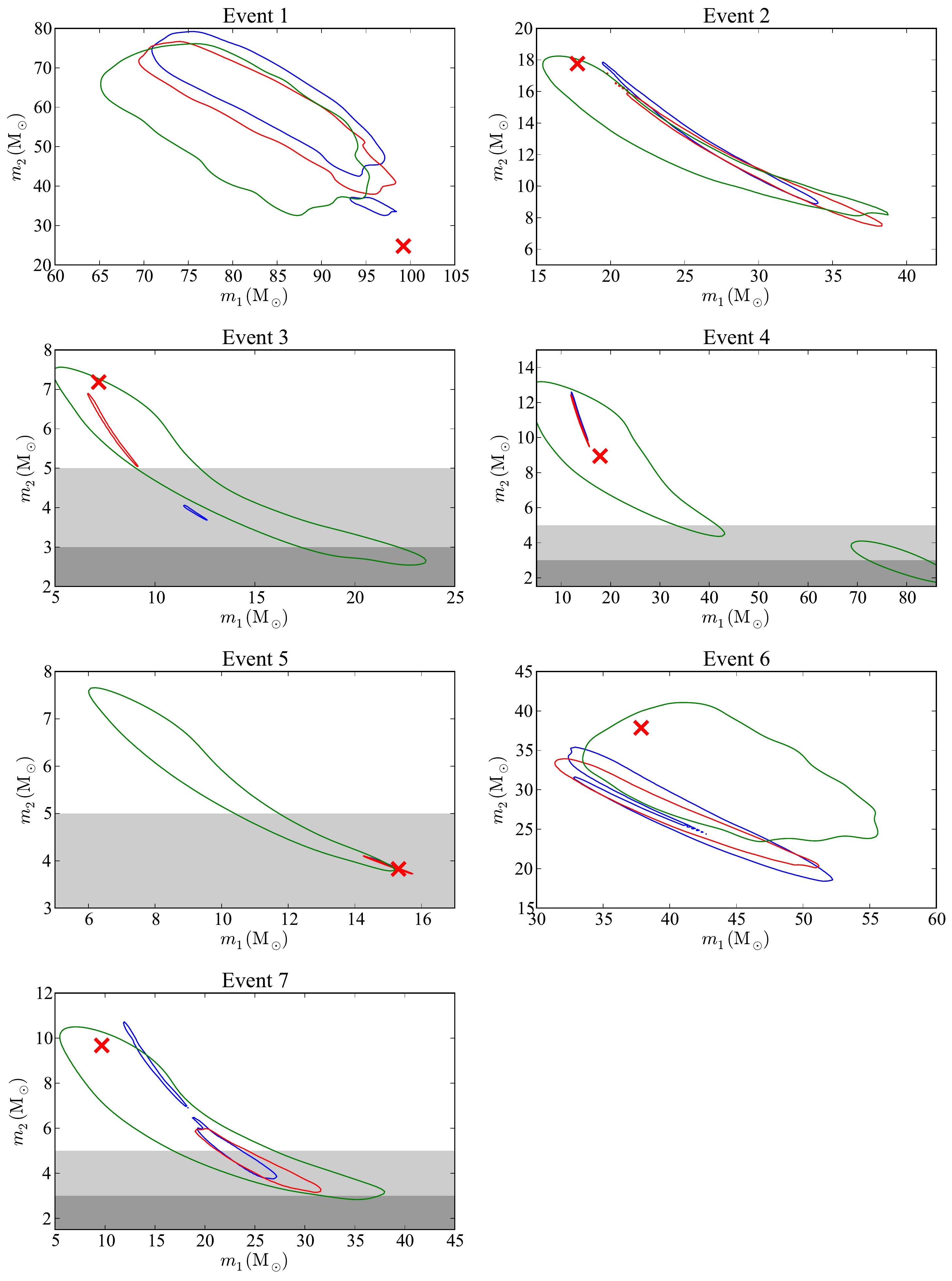}
  \caption{\label{fig:PE_comp_mass} The 95\% credible regions for the \eob
      (blue), \imrns (red), and \imr (green) models.  Injected values are
      indicated by the ``x'' and the neutron star and mass gap regions are
      indicated where relevant.  A strong degeneracy between spin and mass
      ratio results in systematic biases and artificially strong constraints on
      mass estimates when spin is ignored (i.e. \eob, \imrns).  By accounting
      for spin the \imr model produces estimates consistent with the injected
      sources.}
\end{figure}

\begin{figure}[t]
  \includegraphics[width=0.9\textwidth]{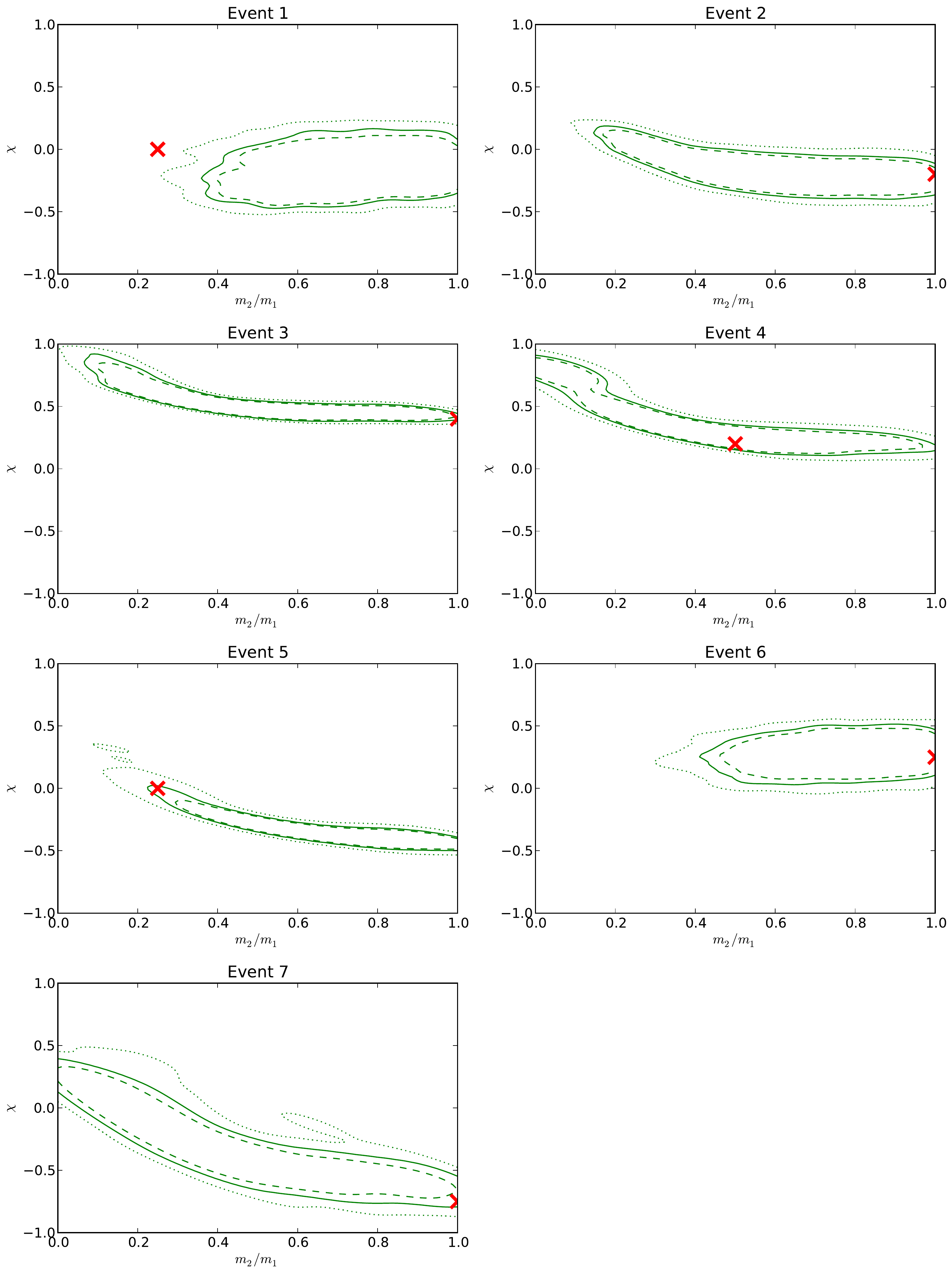} 
  \caption{\label{fig:PE_q_chi} The 90\% (dashed), 95\% (solid), and 99\%
      (dotted) credible regions of the two-dimensional marginalized posterior
      for the mass ratio and effective spin $\chi$ using the \imr model.  The
      strong correlation between mass ratio and spin is responsible for the
      systematically weaker constraints placed on component masses when
      analyzing signals with an aligned-spin model.}
\end{figure}

Figure~\ref{fig:PE_comp_mass} shows the 95\% credible region of the
marginalized posterior in component mass space for the non-spinning and
spinning models.  The credible region is systematically larger for the spinning
model due to the strong degeneracy between the mass ratio and spin magnitude
for aligned-spin models~\cite{Baird:2012cu,Hannam:2013uu}, illustrated in
figure~\ref{fig:PE_q_chi}.

In a real detection scenario, we can never be certain that a source contains
only non-spinning components.  Therefore to evaluate the ability to distinguish
between `typical' stellar-mass black holes and those occupying the mass gap
($\sim 3 - 5 \msun$), we must restrict our attention to the spinning \imr
model.  Event 3 is a spinning system where both components have a mass of 7.18 
\msun.
Despite being well above the mass gap, the degeneracy of the spinning model
results in constraining the lower-mass component to be outside of the mass gap
with only 46\% confidence.  Event 5, on the other hand, contains a lower-mass
component well inside the mass gap, with a mass of 3.83 \msun.  Knowing a
priori that this is a non-spinning system, we would be able to constrain the
mass to be within the gap with 100\% certainty.  However, when including spin,
the \imr model only constrains it to be within the gap with 21\% certainty. The
estimates of the masses and spin are summarized in
Table~\ref{intrinsicPEsummary}.  These results highlight the need to take spin
into account when making any statements about compact object mass from GW data.
We note that spin-aligned systems are the most extreme case of this degeneracy;
if spins are mis-aligned with respect to the binary angular momentum the binary
precesses, which causes phase and amplitude modulations in the observed
waveform. This effect provides additional information to break the degeneracy
between masses and spins that may result in tighter constraints on the
component masses. However, we expect the exact details to depend on the actual
parameters of the observed systems (masses, spins and orientation of the
angular momenta), and studies are ongoing to address this issue.

\begin{figure}[t]
  \subfigure[]{
    \includegraphics[width=0.5\linewidth]{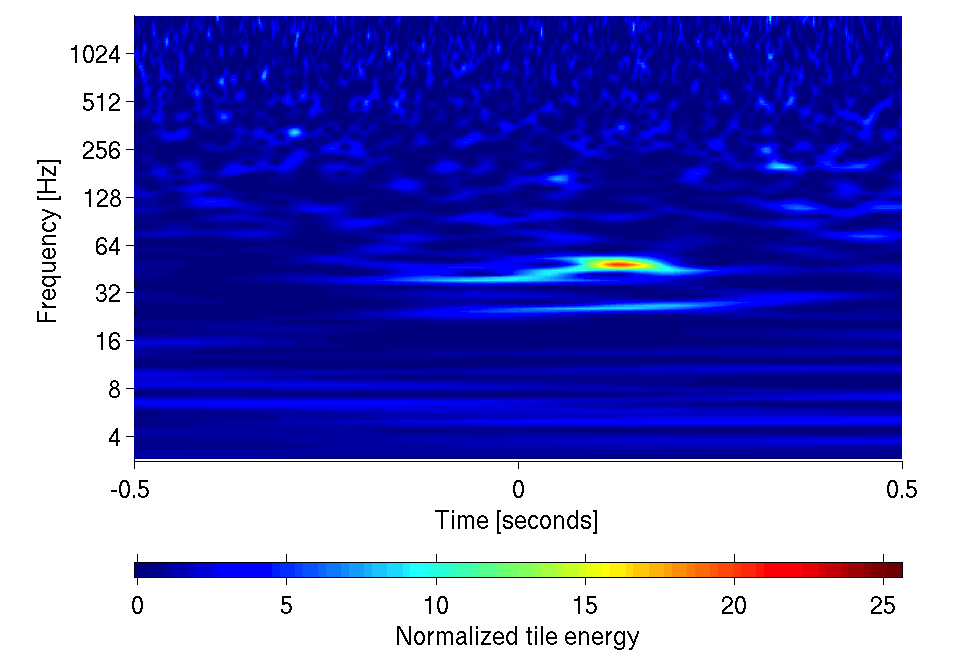}
    \label{subfig:omega}
  }
  \subfigure[]{
    \includegraphics[width=0.48\linewidth]{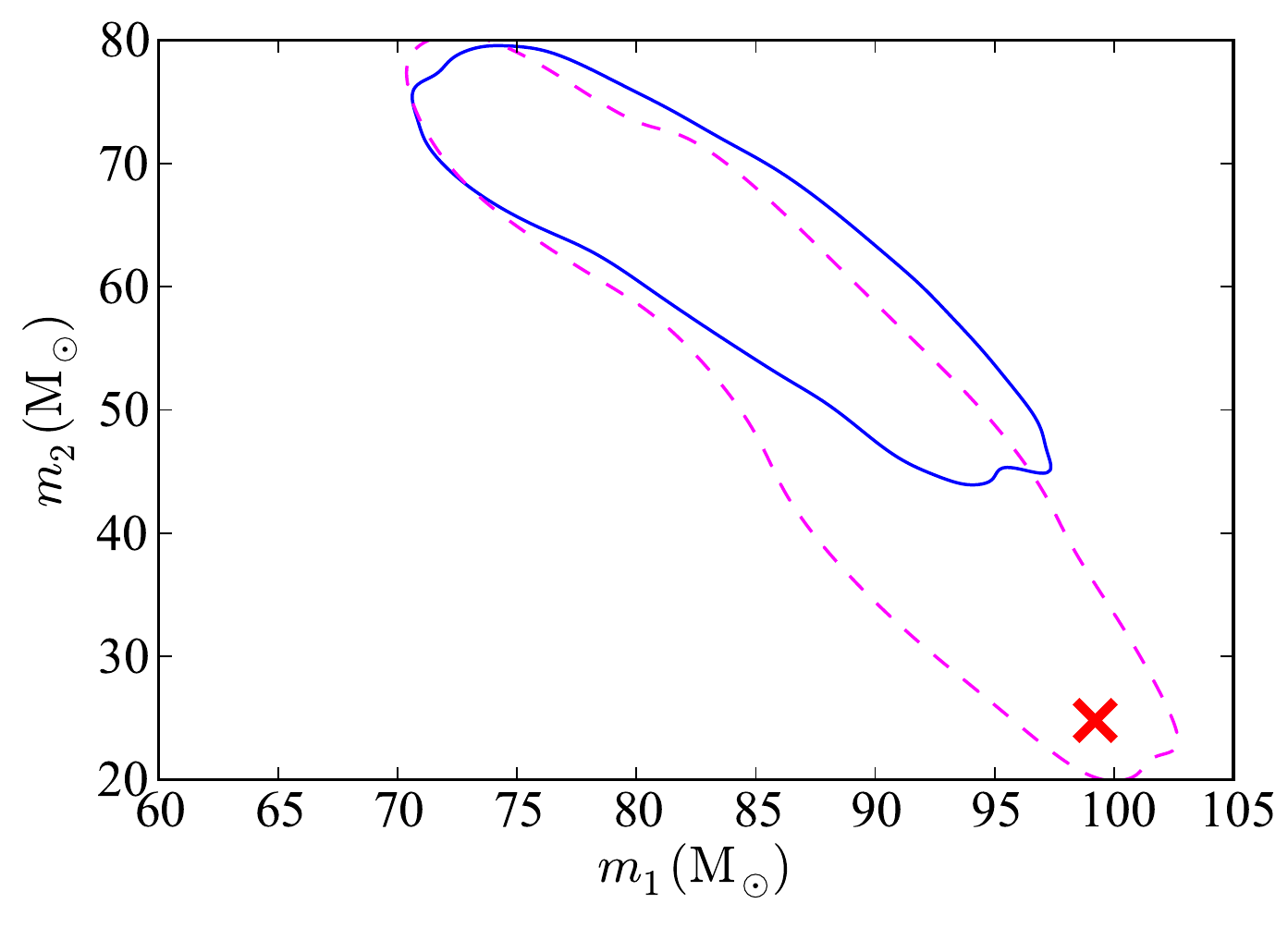}
    \label{subfig:event1_0noise}
  }
  \caption{\label{fig:event1bias} 
      \subref{subfig:omega} shows a spectrogram of the noise in Hanford
      centered on the end time of the injection (before the blind injection 
      is added).  A non-Gaussian feature is present with peak energy slightly 
      after the end time of
      the injection. \subref{subfig:event1_0noise} shows the 95\% credible
      regions for the \eob analysis of the event 1 injection in real noise
      (solid) and noiseless data (dashed).  Injected values are indicated by
      the ``x''.  The fact that the injected values are well outside of the
      95\% credible region when real noise is present, and not in noiseless
      data, leads to the conclusion the non-Gaussian feature in the noise led
      to significant systematic biases in parameter estimates.}
\end{figure}

If the waveform and noise models exactly describe the data, then these Bayesian
credible intervals would be equivalent to frequentist confidence intervals,
meaning that the true parameters would fall within the 95\% credible interval,
for example, for 95\% of injections.  Any errors in the models however,
will introduce systematic biases that break this equivalence.  In these
analyses such biases could be introduced both by the use of model waveforms
that only approximate those that were injected, and by our assumption that the
noise is purely Gaussian.  Since the noise used for this study is real
(recolored) noise recorded by the initial LIGO and Virgo detectors, the
non-Gaussianities inherent with real noise can introduce bias to the parameter
estimates~\cite{Aasi:2013jjl}.  Such biases are most apparent in the first
event.  This non-spinning injection was loud enough that systematic
uncertainties between the numerical relativity and \eob waveforms should be
less than the statistical uncertainty~\cite{Littenberg:2012uj}.  Omega
scan~\cite{Isogai:2010zz} spectrograms show there to be a significant glitch in
the Hanford detector at the time of the injection (see
Fig.~\ref{subfig:omega}).  An additional MCMC analysis was performed on the same
injection made into noiseless data, using the same PSD as estimated for the
event 1 analysis.  Figure~\ref{subfig:event1_0noise} shows the 95\% credible
region of this analysis to indeed constrain the injected values, leading to the
conclusion that the non-Gaussian features of the detector noise led to
significant systematic biases.  Such biases, to varying degrees, are likely for
any signal in noise that is not properly modeled.  It will be crucial for
better noise models to be implemented before the first detections in order to
avoid these biases in parameter estimates.

\begin{figure}[t]
  \includegraphics[scale=.6]{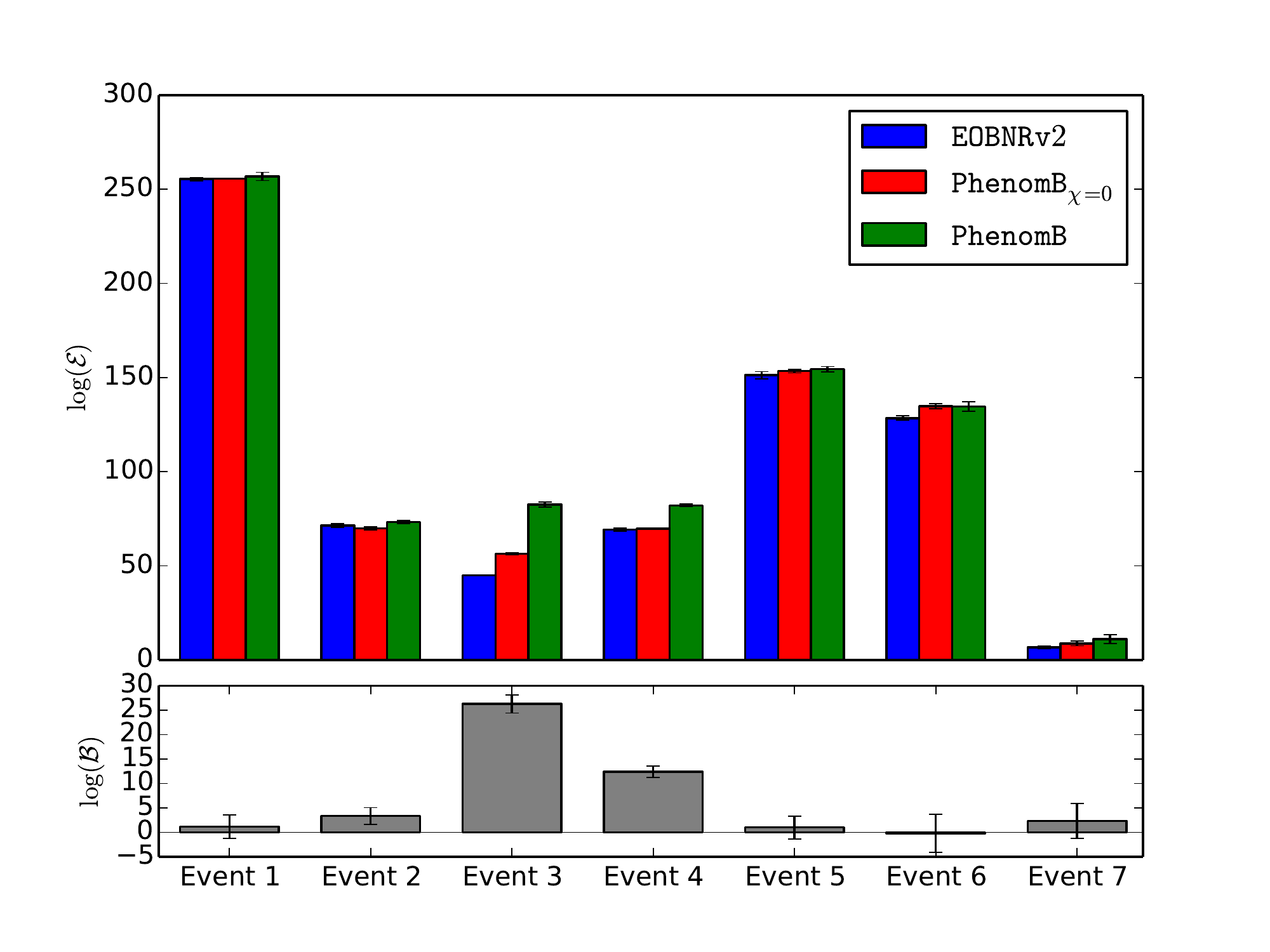}
  \caption{\label{fig:PE_modelComp} (Top) Evidences for the non-spinning (\eob
     and \imrns) and spin-aligned (\imr) models. (Bottom) Bayes factors showing
     the support for the spin-aligned model over the non-spinning \imrns model.
     Events 3 and 4 are found to be spinning with high certainty.  Event 7, with
     the largest spin magnitude, has little support for spin, though this 
     is likely due to its low SNR.}
\end{figure}

The top of figure~\ref{fig:PE_modelComp} shows the evidence for each model for
the events analyzed.  From the non-spinning (\imrns) and spinning (\imr)
evidences, the support for the presence of spin in each signal can be
quantified using the Bayes factor, defined in equation~(\ref{bayesFactor}).  
The bottom of figure~\ref{fig:PE_modelComp} shows the Bayes factor and 
associated error estimates.

These model comparisons show strong support for the presence of spin for events
3 ($\chi_\text{inj}=0.4$) and 4 ($\chi_\text{inj}=0.20$).  Event 7 had the
greatest spin magnitude at $\chi_\text{inj}=-0.75$, however analyses showed
little evidence of it.  This can likely be attributed to the low network SNR of
event 7 (see table \ref{tab:blindsigs}), which was partly due to the faster
phase evolution of systems with spins counter-aligned to the orbital angular
momentum~\cite{Campanelli:2006uy}.

There has been much work done to quantify the ability of ground-based detectors
to localize GW sources on the
sky~\cite{Veitch:2012df,Aasi:2013wya,Nissanke:2011ax,Fairhurst:2010is}.  The
accuracy of such localizations will be important for triggering electromagnetic
(EM) followup of GW detections.  Though most of this sky-localization work has 
focused on binary
neutron star mergers due to their likely association with short gamma ray
bursts~\cite{Eichler:1989ve,Narayan:1992iy,Fong:2013iia} and numerous other
proposed emission mechanisms (e.g.  r-process~\cite{Rosswog:2000qm}, etc.),
there is still much to be learned from accurate binary black hole merger
localization.  Since such mergers have never been observed, the possibility of
an unexpected EM emission mechanism warrants the EM followup of the first
detections.

\begin{figure}[t]
  \includegraphics[width=0.9\textwidth]{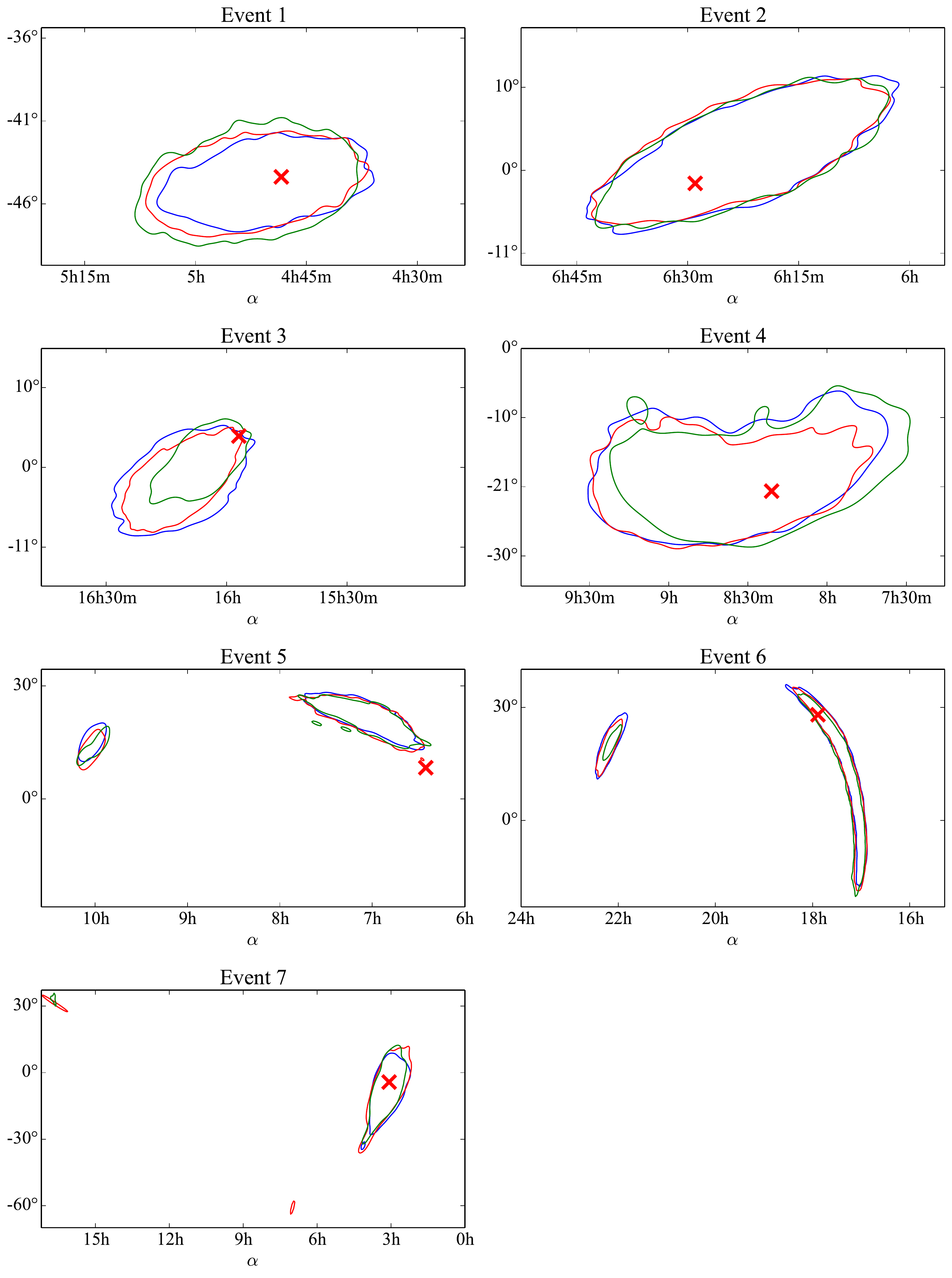}
  \caption{\label{fig:PE_skyLoc} 95\% credible regions for the sky position
      of all seven events, using the \eob (blue), \imrns (red), and \imr (green)
      models.  Despite the substantial differences between these models,
      their sky localization ability is very consistent.}
\end{figure}

Figure~\ref{fig:PE_skyLoc} shows the two dimensional marginalized distribution
for the sky position of each event.  Estimates of each sky position and the
areas of the 95\% credible regions are given in Table \ref{extrinsicPEsummary}.
The constraints on sky position are remarkably consistent across the models
used.  This is of great importance for low latency sky localization efforts,
where time is of the essence.  These results suggest that the more physically
accurate, and computationally expensive, waveform models do not provide
significantly more precise or accurate estimates of source sky positions.

\begin{table}
  \caption{\label{intrinsicPEsummary}
      Summary of the mass and spin estimates for events 1-7 for each of the 
models used.  Median values are
      quoted along with the upper and lower bounds of the 95\% credible 
intervals.  The injected values for each
      event are included for reference.}
  \begin{tabular}{|l|l||c|c||c|c||c|c||}
    \hline Event & \multirow{2}{*}{Model} 
&\multicolumn{2}{c||}{$m_1\,(\mathrm{M}_\odot)$} & 
\multicolumn{2}{c||}{$m_2\,(\mathrm{M}_\odot)$} & \multicolumn{2}{c||}{$\chi$} 
\\
    \cline{3-8} 
 ID & & Median & True & Median & True & Median & True \\
    \hline \hline
  \multirow{3}{*}{1} & \imr & $79.5^{68.6}_{93.6}$ & \multirow{3}{*}{$99.2$} & 
$58.2^{36.7}_{71.6}$ & \multirow{3}{*}{$24.8$} & $-0.184^{-0.417}_{0.128}$ & 
\multirow{3}{*}{$0$} \\
    \cline{2-2}\cline{3-3} \cline{5-5} \cline{7-7}
& \imrns & $81.3^{72.6}_{96.1}$ & & $63.1^{40.2}_{72.7}$ & & - & \\
    \cline{2-2}\cline{3-3} \cline{5-5} \cline{7-7}
& \eob & $83.2^{74.1}_{96}$ & & $64.3^{37.7}_{74.8}$ & & - & \\
    \hline \hline
  \multirow{3}{*}{2} & \imr & $22.1^{17.4}_{37.8}$ & \multirow{3}{*}{$17.8$} & 
$13.6^{8.43}_{17.3}$ & \multirow{3}{*}{$17.8$} & $-0.18^{-0.37}_{0.0999}$ & 
\multirow{3}{*}{$-0.2$} \\
    \cline{2-2}\cline{3-3} \cline{5-5} \cline{7-7}
& \imrns & $32.2^{20.1}_{36.8}$ & & $9.82^{8.15}_{16.8}$ & & - & \\
    \cline{2-2}\cline{3-3} \cline{5-5} \cline{7-7}
& \eob & $27.9^{19.9}_{33.2}$ & & $11.8^{9.32}_{17.5}$ & & - & \\
    \hline \hline
  \multirow{3}{*}{3} & \imr & $11^{7.32}_{26.2}$ & \multirow{3}{*}{$7.18$} & 
$4.82^{2.55}_{7.04}$ & \multirow{3}{*}{$7.18$} & $0.497^{0.415}_{0.923}$ & 
\multirow{3}{*}{$0.4$} \\
    \cline{2-2}\cline{3-3} \cline{5-5} \cline{7-7}
& \imrns & $7.64^{6.83}_{9.16}$ & & $6.04^{5.04}_{6.75}$ & & - & \\
    \cline{2-2}\cline{3-3} \cline{5-5} \cline{7-7}
& \eob & $11.9^{11.5}_{12.6}$ & & $3.91^{3.7}_{4.03}$ & & - & \\
    \hline \hline
  \multirow{3}{*}{4} & \imr & $19.8^{12.9}_{82.6}$ & \multirow{3}{*}{$17.9$} & 
$8.26^{2.78}_{12.4}$ & \multirow{3}{*}{$8.95$} & $0.294^{0.161}_{0.862}$ & 
\multirow{3}{*}{$0.2$} \\
    \cline{2-2}\cline{3-3} \cline{5-5} \cline{7-7}
& \imrns & $13.4^{12.3}_{15.8}$ & & $11.2^{9.46}_{12.2}$ & & - & \\
    \cline{2-2}\cline{3-3} \cline{5-5} \cline{7-7}
& \eob & $13.3^{12.4}_{15.4}$ & & $11.5^{9.89}_{12.3}$ & & - & \\
    \hline \hline
  \multirow{3}{*}{5} & \imr & $9.12^{7.38}_{16.9}$ & \multirow{3}{*}{$15.3$} & 
$5.9^{3.57}_{7.22}$ & \multirow{3}{*}{$3.83$} & $-0.36^{-0.434}_{0.1}$ & 
\multirow{3}{*}{$0$} \\
    \cline{2-2}\cline{3-3} \cline{5-5} \cline{7-7}
& \imrns & $15.1^{14.3}_{15.6}$ & & $3.89^{3.75}_{4.09}$ & & - & \\
    \cline{2-2}\cline{3-3} \cline{5-5} \cline{7-7}
& \eob & $15^{14.5}_{15.4}$ & & $3.92^{3.81}_{4.05}$ & & - & \\
    \hline \hline
  \multirow{3}{*}{6} & \imr & $42.4^{35.9}_{54.4}$ & \multirow{3}{*}{$37.9$} & 
$32.5^{24.4}_{38.9}$ & \multirow{3}{*}{$37.9$} & $0.277^{0.0582}_{0.468}$ & 
\multirow{3}{*}{$0.25$} \\
    \cline{2-2}\cline{3-3} \cline{5-5} \cline{7-7}
& \imrns & $39.2^{33.1}_{49.7}$ & & $27.5^{21}_{32.7}$ & & - & \\
    \cline{2-2}\cline{3-3} \cline{5-5} \cline{7-7}
& \eob & $40.1^{33.9}_{51}$ & & $27.7^{19.5}_{33.7}$ & & - & \\
    \hline \hline
  \multirow{3}{*}{7} & \imr & $16.3^{10.1}_{39.8}$ & \multirow{3}{*}{$9.67$} & 
$6.27^{3.18}_{9.61}$ & \multirow{3}{*}{$9.67$} & $-0.332^{-0.673}_{0.286}$ & 
\multirow{3}{*}{$-0.75$} \\
    \cline{2-2}\cline{3-3} \cline{5-5} \cline{7-7}
& \imrns & $24.8^{18.1}_{29.7}$ & & $4.45^{3.61}_{7.17}$ & & - & \\
    \cline{2-2}\cline{3-3} \cline{5-5} \cline{7-7}
& \eob & $23.1^{13.3}_{25.7}$ & & $4.8^{4.22}_{9.62}$ & & - & \\
    \hline 
  \end{tabular}
\end{table}

\begin{table}
  \caption{\label{extrinsicPEsummary}
      Summary of the sky location estimates for events 1-7 for each of the 
models used.  Median values are
      quoted along with the upper and lower bounds of the 95\% credible 
intervals, as well as the area of the
      95\% credible regions in the sky.  The injected values for each event are 
included for reference.}
  \begin{tabular}{|l|l||c|c||c|c||c|}
    \hline Event & \multirow{2}{*}{Model} &\multicolumn{2}{c||}{$\alpha$} & 
\multicolumn{2}{c||}{$\delta$} & Sky Area \\
    \cline{3-6} 
 ID & & Median & True & Median & True & (sq. deg.) \\
    \hline \hline
  \multirow{3}{*}{1} & \imr & $1.28^{1.23}_{1.33}$ & \multirow{3}{*}{$1.26$} & 
$-0.767^{-0.807}_{-0.684}$ & \multirow{3}{*}{$-0.757$} & 32.4 \\
    \cline{2-2}\cline{3-3} \cline{5-5}\cline{7-7}
& \imrns & $1.27^{1.22}_{1.32}$ & & $-0.764^{-0.803}_{-0.712}$ & & 26.6 \\
    \cline{2-2}\cline{3-3} \cline{5-5}\cline{7-7}
& \eob & $1.26^{1.21}_{1.32}$ & & $-0.763^{-0.801}_{-0.722}$ & & 22.5 \\
    \hline \hline
  \multirow{3}{*}{2} & \imr & $1.67^{1.6}_{1.74}$ & \multirow{3}{*}{$1.7$} & 
$0.0402^{-0.0877}_{0.163}$ & \multirow{3}{*}{$-0.0276$} & 103 \\
    \cline{2-2}\cline{3-3} \cline{5-5}\cline{7-7}
& \imrns & $1.67^{1.6}_{1.74}$ & & $0.0453^{-0.0827}_{0.163}$ & & 100 \\
    \cline{2-2}\cline{3-3} \cline{5-5}\cline{7-7}
& \eob & $1.67^{1.6}_{1.74}$ & & $0.0406^{-0.0921}_{0.164}$ & & 100 \\
    \hline \hline
  \multirow{3}{*}{3} & \imr & $4.21^{4.17}_{4.26}$ & \multirow{3}{*}{$4.18$} & 
$0.00761^{-0.0587}_{0.086}$ & \multirow{3}{*}{$0.0684$} & 41.7 \\
    \cline{2-2}\cline{3-3} \cline{5-5}\cline{7-7}
& \imrns & $4.24^{4.18}_{4.29}$ & & $-0.0377^{-0.11}_{0.0776}$ & & 65.7 \\
    \cline{2-2}\cline{3-3} \cline{5-5}\cline{7-7}
& \eob & $4.24^{4.17}_{4.3}$ & & $-0.0388^{-0.126}_{0.0761}$ & & 80.0 \\
    \hline \hline
  \multirow{3}{*}{4} & \imr & $2.19^{2.01}_{2.41}$ & \multirow{3}{*}{$2.19$} & 
$-0.333^{-0.441}_{-0.132}$ & \multirow{3}{*}{$-0.36$} & 430 \\
    \cline{2-2}\cline{3-3} \cline{5-5}\cline{7-7}
& \imrns & $2.3^{2.08}_{2.45}$ & & $-0.36^{-0.459}_{-0.197}$ & & 301 \\
    \cline{2-2}\cline{3-3} \cline{5-5}\cline{7-7}
& \eob & $2.3^{2.06}_{2.45}$ & & $-0.344^{-0.451}_{-0.14}$ & & 470 \\
    \hline \hline
  \multirow{3}{*}{5} & \imr & $1.84^{1.72}_{2.6}$ & \multirow{3}{*}{$1.68$} & 
$0.376^{0.193}_{0.44}$ & \multirow{3}{*}{$0.144$} & 215 \\
    \cline{2-2}\cline{3-3} \cline{5-5}\cline{7-7}
& \imrns & $1.85^{1.74}_{2.62}$ & & $0.377^{0.209}_{0.45}$ & & 199 \\
    \cline{2-2}\cline{3-3} \cline{5-5}\cline{7-7}
& \eob & $1.86^{1.74}_{2.63}$ & & $0.389^{0.223}_{0.455}$ & & 150 \\
    \hline \hline
  \multirow{3}{*}{6} & \imr & $4.5^{4.44}_{5.78}$ & \multirow{3}{*}{$4.68$} & 
$0.141^{-0.247}_{0.521}$ & \multirow{3}{*}{$0.49$} & 207 \\
    \cline{2-2}\cline{3-3} \cline{5-5}\cline{7-7}
& \imrns & $4.52^{4.44}_{5.84}$ & & $0.233^{-0.241}_{0.542}$ & & 294 \\
    \cline{2-2}\cline{3-3} \cline{5-5}\cline{7-7}
& \eob & $4.56^{4.44}_{5.85}$ & & $0.309^{-0.192}_{0.551}$ & & 342 \\
    \hline \hline
  \multirow{3}{*}{7} & \imr & $0.83^{0.684}_{1.67}$ & \multirow{3}{*}{$0.806$} & 
$-0.121^{-0.532}_{0.159}$ & \multirow{3}{*}{$-0.0736$} & 551 \\
    \cline{2-2}\cline{3-3} \cline{5-5}\cline{7-7}
& \imrns & $0.836^{0.662}_{1.84}$ & & $-0.117^{-0.595}_{0.17}$ & & 716 \\
    \cline{2-2}\cline{3-3} \cline{5-5}\cline{7-7}
& \eob & $0.836^{0.653}_{1.02}$ & & $-0.135^{-0.495}_{0.0922}$ & & 627 \\
    \hline 
  \end{tabular}
\end{table}

\section{Sensitivity evaluation}
\label{sec:sensitivity}

A large set of simulated signals, distributed as described in section 
\ref{sec:parameters}, was used to assess the sensitivity of the 
pipelines to observe numerical relativity signals buried in data taken from 
Initial LIGO and Initial Virgo and recoloured to predicted early advanced 
detector observing runs, as described in section~\ref{sec:noise}. Here, we use 
the CBC search 
pipelines (low-mass and high-mass \ihope{}) to assess search 
efficiency in the chosen mass region: total mass between 10 and 
100 solar masses.

Each signal in these large simulation sets was added at a random time
to the 2-month period of recolored data. Coalescence times were limited to 
times where at least two observatories were operating and no data-quality flags 
were active. We then search for each
signal using the \ihope{} pipeline, as described in section
\ref{sec:pipelines}. In the plots that follow we treat a simulated
signal as ``detected'' if there was no louder background event in the
100 time-slide trials that were performed to estimate the search
background. As two \ihope{} searches are performed, ``low-mass'' and
``high-mass'', we treat simulated signals as detected if \emph{either}
search recovers the signal with more significance than its
corresponding 100 background trials.

\begin{figure}
\centering
\includegraphics[width=0.495\textwidth]
{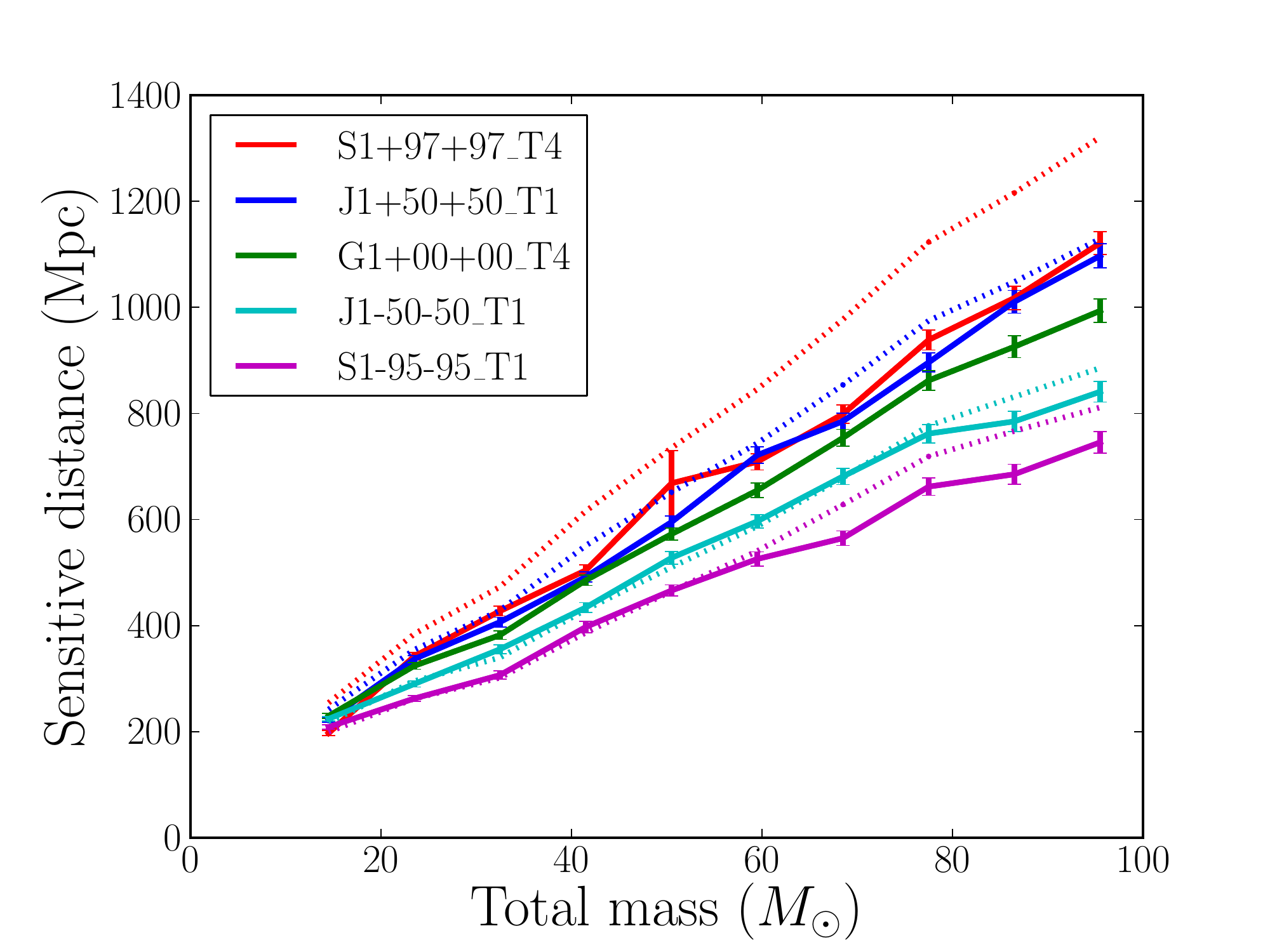}
\includegraphics[width=0.495\textwidth]
{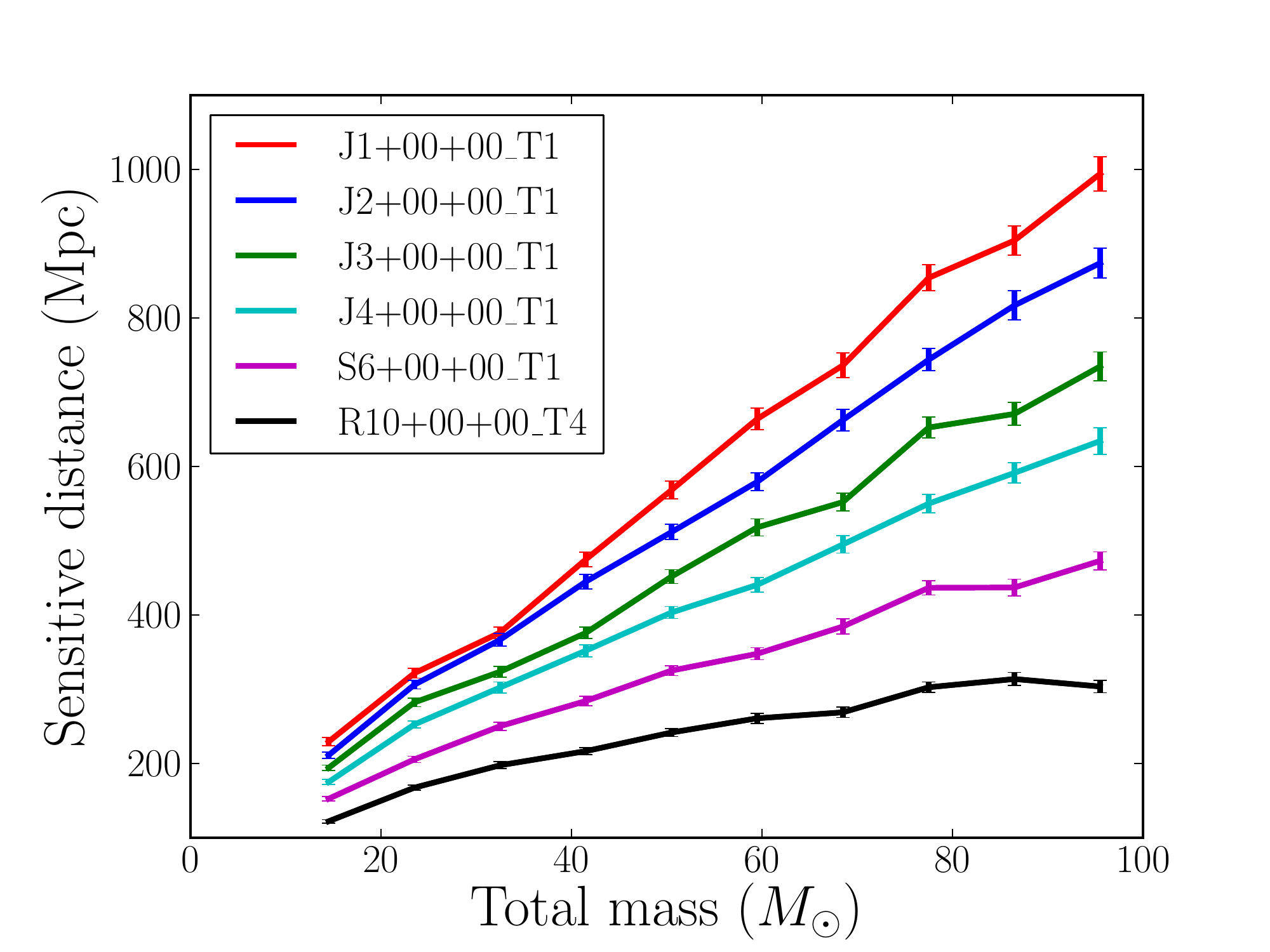}
\includegraphics[width=0.495\textwidth]
{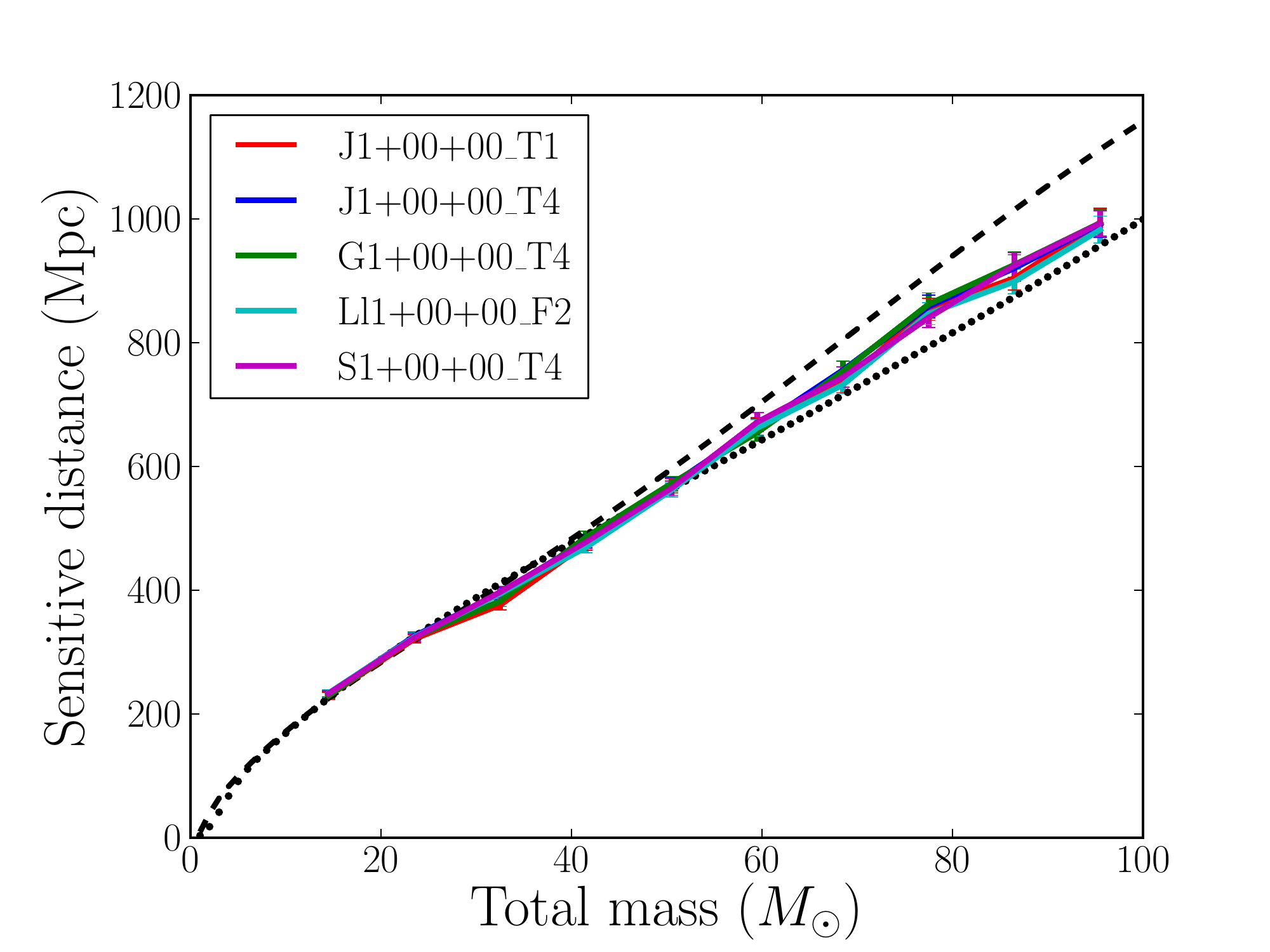}
\includegraphics[width=0.495\textwidth]
{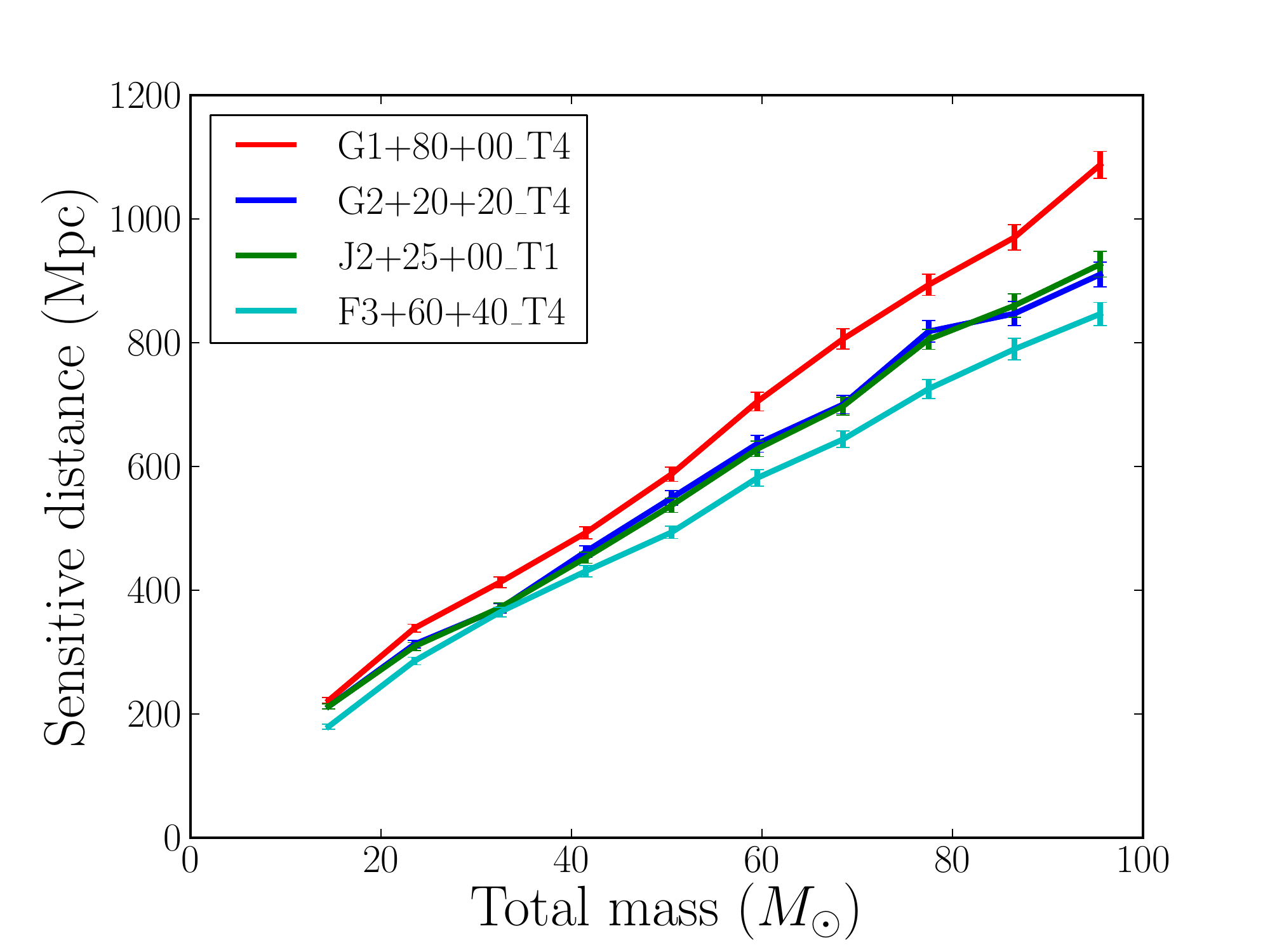}
\caption{\label{fig:sens_dists_1}
Volume-weighted average sensitive distance as a function of total mass, for a 
variety of NR waveforms. The sensitive distance is defined as the 
volume-weighted average distance 
up to which a CBC signal with a given waveform and total mass can be detected, 
where we average over the source extrinsic parameters and over the varying 
detector noise levels in the data set via Monte Carlo integration.
\emph{Top left}---Waveforms with equal component masses and various spins. Here 
the dotted lines represent predicted performances if an aligned-spin search had 
been performed. Details of this prediction are given in section 
\ref{sec:sensitivity}.
\emph{Top right}---Waveforms with zero component spins and various mass ratios.
\emph{Bottom left}---Waveforms with equal-mass, non-spinning components. The 
dashed black line represents the predicted sensitivity curve of non-spinning 
waveforms with the aLIGO sensitivity curve, as described in 
\ref{sec:sensitivity}. The dotted black line represents the predicted 
sensitivity with tha AdV sensitivity curve.
\emph{Bottom right}---Waveforms with unequal spins, or unequal component masses 
and non-zero spins.
The waveform designations are given in Table 
\ref{tab:ninja2_submissions1}. These plots were generated using the data 
described in section~\ref{sec:noise} and the distribution of signals described 
in section~\ref{sec:parameters}.}
\end{figure}

The results of the injection campaign are summarized in 
figure~\ref{fig:sens_dists_1}, which shows the dependence of the search
sensitive distance on the chosen NR waveform and the total mass value. 
The sensitive distance is defined as the volume-weighted average distance 
up to which a CBC signal with a given waveform and total mass can be detected, 
where we average over the source extrinsic parameters and over the varying 
detector noise levels in the data set via Monte Carlo integration. 
In order to obtain a clear comparison between different NR waveforms, we used
\emph{the same set of random parameters} (including total masses, coalescence
times and orientation angles) for each set of $\sim$24000 injected signals, thus
statistical errors will tend to be correlated between different waveforms.
 
In figure~\ref{fig:sens_dists_1}, top left, we plot the sensitive 
distance as a function of total mass for a number of equal mass NR waveforms 
with varying component spin magnitudes. The two waveforms with anti-aligned 
spins have smaller
sensitive distances than the other waveforms (see also 
\cite{Aasi:2012rja}). This is expected; systems with 
anti-aligned spin emit less power in gravitational waves than non-spinning 
systems \cite{Buonanno:2005xu,Ajith:2009bn} and also may not match well with 
the template bank 
of non-spinning waveforms that was used \cite{Ajith:2012mn}. At the largest 
masses in this study we can detect the two 
aligned-spin waveforms at distance $\sim10-15\%$ larger than the non-spinning 
waveform. 
However, at the smallest masses used, the non-spinning waveform can be seen to 
comparable distances. Again this is a combination of two factors. Aligned-spin 
waveforms emit more power in gravitational waves than non-spinning systems, but 
the bank of non-spinning waveforms will not capture all of that power.
This emphasizes that investigating the use of template waveforms incorporating 
spin effects is a necessity for the advanced detector era. 
To illustrate this further the dotted lines in this figure represent a 
prediction of the sensitivity that would have been obtained if a bank of 
aligned-spin template waveforms had been used. This prediction was obtained for 
the (anti)aligned-spin waveforms by multiplying the ratio of the expected 
signal-to-noise ratio of each spinning waveform and the non-spinning 
waveform with the performance of the non-spinning waveform. This is described by
\begin{equation}
 D_{\mathrm{sens}}^i (M) = \frac{\sigma^i(M)}{\sigma^{\mathrm{NS}}(M)}
D_{\mathrm{sens}}^{\mathrm{NS}} (M).
\end{equation}
Here $i$ is used to denote the aligned-spin waveform for which the sensitive 
distance, $D_{\mathrm{sens}}^i$, is to be predicted. $\mathrm{NS}$ is used to 
denote the non-spinning waveform. $\sigma$ is a measure of the signal power 
given by the inner product defined in equation~(\ref{eq:InnerProduct})
\begin{equation}
   \sigma^i = \InnerProduct{h^i|h^i}^{1/2},
\end{equation}
where $h$ denotes the waveform corresponding to $i$. We can see from the plot 
that the distance sensitivity to spinning waveforms using a non-spinning bank 
is not as large as the predicted values from using (anti)aligned-spin template 
banks. 
This is especially true for the highly aligned-spin waveform where the 
predicted 
sensitive distance is $\sim 15\%$ larger than the obtained distance for systems 
with total mass $> 40M_{\odot}$. We note that the actual sensitivity 
improvement of a search using (anti)aligned-spin templates may not be as much 
as predicted here because such a search would require a larger number of 
templates and therefore provide more chances to obtain large SNR when 
matched-filtered against the underlying noise. Therefore the detection 
threshold 
when performing an aligned-spin search will increase with respect to a 
search using a non-spinning template bank. However, even a factor of 10 
increase in the number of \emph{independent} templates will only increase the 
expected SNR of the loudest background event by less than $5\%$, if 
Gaussian noise is assumed~\cite{Harry:2013tca}. Work is ongoing to assess how 
much larger template banks of aligned-spin BBH waveforms are when compared to 
banks restricted to only non-spinning waveforms and to accurately compare the 
performance of such searches.

In the top right panel of figure~\ref{fig:sens_dists_1} we plot the horizon 
distance as a function of total mass for a number of non-spinning NR waveforms 
with varying mass ratio. The sensitivity to these systems, for the same total 
mass, increases as the mass ratio (which we take to be greater than or equal to
$1$) decreases. This is because, to leading order, 
the gravitational-wave power emitted during the inspiral phase is dependent on 
the chirp mass \cite{Peters:1963ux}; for the same total mass, a system with 
higher mass ratio will have a smaller chirp mass.

The bottom left panel of figure~\ref{fig:sens_dists_1} shows the
horizon distance as a function of total mass for the 5 non-spinning,
equal-mass waveforms that were submitted. As expected, there is no
significant discrepancy between these waveforms. We remind the reader that the 
same set of source parameters was used for each of these 5 sets of injections. 
Therefore statistical errors are strongly correlated between the results for
different waveforms. The dashed and dotted black lines on this plot represent 
a prediction of the sensitive distance to non-spinning signals for the 
early aLIGO and early AdV noise curves respectively. This prediction was 
obtained by calculating the horizon distance, defined in section 
\ref{sec:noise}, for the $G1+00+00\_T4$ waveform for both the early aLIGO and 
early AdV sensitivity curves as a function of mass. This measurement is then 
rescaled by a factor of 2.26 to account for the fact that the observatories do 
not have equal power to all orientations and sky locations~\cite{Finn:1992xs}. 
We note that the obtained results agree well with this prediction, and fall 
between the early aLIGO and early AdV predictions when the two diverge. This 
is expected, as detection in \ihope{} is dominated by the sensitivity of the 
second most sensitive detector operating at the time~\cite{Babak:2012zx}. For 
times when at least two observatories were operating, the NINJA-2 dataset 
approximately consists of 50\% of time when only one of the LIGO detectors and 
Virgo were operating and 50\% of time when both 
LIGO detectors were operating, including time when Virgo is operating and when 
it is not. As the early aLIGO sensitivity does not drop below the early AdV 
sensitivity for the mass range considered, we expect the obtained sensitivity 
curve to lie roughly in the middle of the two predictions, and this is what we 
observe. It is worth pointing out that the \ihope{} search is able to acheive 
this Gaussian-noise-predicted sensitivity, even though this analysis is run on 
real data, which includes non-stationarity and non-Gaussian transients.

Finally, in the bottom right panel of figure~\ref{fig:sens_dists_1} we show 
sensitive distances for a number of waveforms with unequal spins, or unequal 
masses and non-zero spins.

\section{Conclusion}
\label{sec:conclusions}

This paper presents the first systematic study to assess the ability to 
detect numerically modelled binary black hole data in real data taken from 
Initial LIGO and Virgo and recolored to predicted sensitivity curves of 
Advanced LIGO and Advanced Virgo in early observing runs. Building upon the 
work of the first NINJA project, this work, the culmination of the second NINJA 
project, studies the ability to do gravitational wave astronomy on a set of 60 
binary black hole hybrid waveforms submitted by 8 numerical relativity groups. 

In this work, a set of 7 numerically modelled binary black hole waveforms were 
added into the recolored data. This data was distributed to analysts with no 
knowledge of the parameters of the systems. The unmodelled gravitational 
waveform search pipeline, cWB, was able to recover one of these signals with an 
estimated false alarm rate of 1 every 47 years. The matched-filtered compact 
binary merger search pipeline, \ihope{}, using a bank of BBH IMR waveforms, 
which were not calibrated against the NR signals used in NINJA-2, was able 
to recover 6 of the waveforms with false alarm rate upper limits ranging 
between 
1 every 300 years and 1 every 2500 years. 

A range of parameter estimation codes were run on the 7 blind injections 
that were added to the data used in this work. Though only results from
the MCMC sampler were shown, these results proved to be statistically
equivalent to estimates produce by the nested sampling and multinest
samplers. These results demonstrate that it will be difficult to produce 
precise estimate of black hole component masses and spins because of intrinsic 
degeneracies between these parameters in the emitted waveforms. For some of the 
BBH blind injections we find that a neutron-star--black-hole coalescence cannot 
be ruled out. We also demonstrate the sensitivity of current parameter 
estimation algorithms to non-Gaussian features in the data and explore the 
ability to perform sky-localization of BBH observations.

A large-scale monte-carlo study was conducted to assess the efficiency of the 
\ihope{} search pipeline as a function of the mass and angular momenta of the 
component black holes. We find that for non-spinning equal mass waveforms the 
sensitivity of the \ihope{} search pipeline in real noise, including 
non-Gaussian artifacts, agrees well with predictions obtained using a 
Gaussian-noise assumption. We have found evidence that adding waveform models 
that include the effects of spin into the search pipeline will increase the 
efficiency of binary black hole observation. We have also demonstrated that the 
ability to recover numerical relativity waveforms, with identical parameters, 
but submitted by different groups, is indistinguishable up to statistical 
errors.

These results represent the next step for the NINJA collaboration;
they address shortcomings in NINJA-1 while paving the way
for future work.  In a sense this paper represents a baseline, as it
measures the ability of current gravitational-wave analyses to detect
and recover the parameters of an important subset of possible BBH
signals in non-Gaussian noise in the advanced detector era. 
From this baseline there are
multiple directions in which NINJA can expand. On the NR front,
groups are continuing to fill in the parameter space.  As shown in
figure~\ref{fig:ParameterSpace}, even within the subspace of systems
with (anti-)aligned spins there are large regions left to explore.
Although NINJA-2 chose not to consider precessing signals many groups
already have or are working on such simulations~\cite{
Campanelli:2008nk, Tichy:2007hk, Tichy:2008du,Schmidt:2010it,
O'Shaughnessy:2011fx,Mroue:2012kv,O'Shaughnessy:2012vm,
O'Shaughnessy:2012ay,Hinder:2013oqa,Taracchini:2013rva}. Similarly, although 
the analyses used only the
$\ell=m=2$ mode in this work, it is expected that higher modes will be
important for detection and parameter
recovery~\cite{Pekowsky:2012sr,Healy:2013jza,McWilliams:2010eq,Brown:2012nn,
Pan:2011gk}.  Additional modes have been provided for many of
the waveforms in the NR catalog, although they have not yet been
validated.  In all cases, as additional waveforms and modes become
available they can be injected into the noise allowing for systematic
tests of both detection and parameter estimation analyses.

In parallel the detection and parameter estimation analyses continue to 
evolve and improve. There is much development work ongoing to improve the 
analytical waveform models that are used in analysis pipelines, particularly 
for inspiral-merger-ringdown waveforms. It seems likely that before the 
first aLIGO and AdV observation runs generic fast IMR precessing analytic 
models will be 
available~\cite{Santamaria:2010yb,Taracchini:2012ig,Pan:2013rra,Hannam:2013oca,
Taracchini:2013rva}.
Improvements in how detection pipelines deal with non-Gaussianities are being 
explored to attempt to achieve the maximum possible sensitivity to BBH signals 
across the parameter space. A number of efforts are ongoing to implement 
aligned-spin waveform models into  search algorithms. As we have demonstrated 
here, this will increase sensitivity to BBH systems with aLIGO and 
AdV~\cite{Brown:2012qf,Ajith:2012mn,Harry:2013tca}.
Work is also underway to develop more realistic models 
of detector noise for parameter estimation pipelines, which account for the 
non-stationarity and non-Gaussianity present in real 
noise~\cite{Littenberg:2013gja}.  Accounting for such features
is expected to greatly reduce systematic biases in the recovered masses and 
spins, such as those seen in event 1.
The results presented here can provide a measure against which these
next-generation analyses can be compared, in a way that measures not only their
response to signals but also to realistic noise.

\section*{Acknowledgements}

The authors gratefully acknowledge the support of the United States
National Science Foundation for the construction and operation of the
LIGO Laboratory, the Science and Technology Facilities Council of the
United Kingdom, the Max-Planck-Society, and the State of
Niedersachsen/Germany for support of the construction and operation of
the GEO600 detector, and the Italian Istituto Nazionale di Fisica
Nucleare and the French Centre National de la Recherche Scientifique
for the construction and operation of the Virgo detector. The authors
also gratefully acknowledge the support of the research by these
agencies and by the Australian Research Council, 
the International Science Linkages program of the Commonwealth of Australia,
the Council of Scientific and Industrial Research of India, 
the Istituto Nazionale di Fisica Nucleare of Italy, 
the Spanish Ministerio de Econom\'ia y Competitividad,
the Conselleria d'Economia Hisenda i Innovaci\'o of the
Govern de les Illes Balears, the Foundation for Fundamental Research
on Matter supported by the Netherlands Organisation for Scientific Research, 
the Polish Ministry of Science and Higher Education, the FOCUS
Programme of Foundation for Polish Science,
the Royal Society, the Scottish Funding Council, the
Scottish Universities Physics Alliance, The National Aeronautics and
Space Administration, 
OTKA of Hungary,
the Lyon Institute of Origins (LIO),
the National Research Foundation of Korea,
Industry Canada and the Province of Ontario through the Ministry of Economic 
Development and Innovation, 
the National Science and Engineering Research Council Canada,
the Carnegie Trust, the Leverhulme Trust, the
David and Lucile Packard Foundation, the Research Corporation, and
the Alfred P. Sloan Foundation.

We gratefully acknowledge support from the National Science Foundation
under NSF grants
PHY-1305730, 
PHY-1212426, 
PHY-1229173, 
AST-1028087, 
DRL-1136221, 
OCI-0832606, 
PHY-0903782, 
PHY-0929114, 
PHY-0969855, 
AST-1002667, 
PHY-0963136, 
PHY-1300903, 
PHY-1305387, 
PHY-1204334, 
PHY-0855315, 
PHY-0969111, 
PHY-1005426, 
PHY-0601459, 
PHY-1068881, 
PHY-1005655, 
PHY-0653443, 
PHY-0855892, 
PHY-0914553, 
PHY-0941417, 
PHY-0903973, 
PHY-0955825, 
by NASA grants 
07-ATFP07-0158, 
NNX11AE11G, 
NNX13AH44G, 
NNX09AF96G, 
NNX09AF97G, 
by Marie Curie Grants of the 7th European Community Framework Programme
FP7-PEOPLE-2011-CIG CBHEO No.~293412,
by the DyBHo-256667 ERC Starting Grant, 
and
MIRG-CT-2007-205005/PHY, 
and Science and Technology Facilities Council grants ST/H008438/1
and ST/I001085/1.
Further funding was provided by
the Sherman Fairchild Foundation, 
NSERC of Canada,  
the Canada Research Chairs Program, 
the Canadian Institute for Advanced Research, 
the Ram{\'o}n y Cajal Programme of the Ministry of Education and Science of 
Spain, 
contracts AYA2010-15709, CSD2007-00042, CSD2009-00064 and FPA2010-16495 of the 
Spanish Ministry of Science and Innovation, 
the German Research Foundation, grant SFB/Transregio 7, 
the German Aerospace Center for LISA Germany, 
and the Grand-in-Aid for Scientific Research (24103006). 
Computations were carried out on Teragrid machines Lonestar, Ranger,
Trestles and Kraken under Teragrid allocations 
TG-PHY060027N, 
TG-MCA99S008, 
TG-PHY090095, 
TG-PHY100051, 
TG-PHY990007N, 
TG-PHY090003,  
TG-MCA08X009. 
Computations were also performed on the clusters 
```HLRB-2'' at LRZ Munich,
``NewHorizons'' and ``Blue Sky'' at RIT (funded by NSF Grant Nos.~PHY-1229173, 
AST-1028087, DMS-0820923 and PHY-0722703), 
``Zwicky'' at Caltech (funded by NSF MRI award PHY-0960291),
``Finis Terrae'' (funded by CESGA-ICTS-2010-200),
``Caesaraugusta'' (funded by BSC Grant Nos.~AECT-2011-2-0006, AECT-2011-3-0007),
``MareNostrum''  (funded by BSC Grant Nos.~AECT-2009-2-0017, AECT-2010-1-0008, 
AECT-2010-2-0013, AECT-2010-3-0010, AECT-2011-1-0015, AECT-2011-2-0012),
``VSC'' in Vienna (funded by FWF grant P22498),
``Force'' at GaTech, 
and on the GPC supercomputer at the SciNet HPC Consortium~\cite{Loken:2010}; 
SciNet is
funded by: the Canada Foundation for Innovation under the auspices of
Compute Canada; the Government of Ontario; Ontario Research Fund -
Research Excellence; and the University of Toronto.

\section*{References}
\bibliography{references}

\end{document}